\documentclass[10pt,journal,compsoc]{IEEEtran}
\ifCLASSOPTIONcompsoc
  \usepackage[nocompress]{cite}
\else
  \usepackage{cite}
\fi
\ifCLASSINFOpdf
\else
\fi
\usepackage{amssymb,amsmath,amsthm,enumerate,threeparttable,bm,hyperref}
\usepackage{caption}
\usepackage{subfigure}
\usepackage{algorithm}
\usepackage{algorithmic}
\usepackage{booktabs}
\usepackage{multirow}
\usepackage{colortbl}
\usepackage{threeparttable}

\newtheorem{theorem}{Theorem}
\newtheorem{definition}{Definition}

\newtheorem{example}{Example}

\usepackage{color}
\def\red#1{\textcolor{red}{#1}}
\def\ie{$i.e.$}
\def\eg{$e.g.$}

\usepackage{lipsum}
\usepackage{enumitem}
\usepackage{xcolor}

\def\revision#1{\textcolor{black}{#1}}

\long\def\comment#1{}
\usepackage{setspace}
\usepackage{amsmath}
\usepackage{wrapfig}
\usepackage[export]{adjustbox}
\usepackage[switch]{lineno}

\usepackage{array}
\newcommand{\tabincell}[2]{\begin{tabular}{@{}c#1@{}}#2\end{tabular}}

\begin{document}

\title{MOVE: Effective and Harmless Ownership Verification via Embedded External Features}

\author{Yiming~Li,
        Linghui~Zhu,
        Xiaojun~Jia,
        Yang Bai,
        Yong~Jiang,
        Shu-Tao~Xia,
        Xiaochun~Cao,
        and Kui Ren

\thanks{Yiming Li was with the State Key Laboratory of Blockchain and Data
Security, Zhejiang University, Hangzhou 310007, China, and was also with Tsinghua Shenzhen International Graduate School, Tsinghua University, Shenzhen 5180, China. He is now with Nanyang Technological University, Singapore 639798 (e-mail: \href{mailto:liyiming.tech@gmail.com}{liyiming.tech@gmail.com}).}
\thanks{Linghui Zhu is with Tsinghua Shenzhen International Graduate School, Tsinghua University, Shenzhen, China (e-mail: \href{zlh20@mails.tsinghua.edu.cn}{zlh20@mails.tsinghua.edu.cn}).}%
\thanks{Xiaojun Jia is with College of Computing and Data Science, Nanyang Technological University, Singapore 639798 (e-mail: \href{jiaxiaojunqaq@gmail.com}{jiaxiaojunqaq@gmail.com}).}
\thanks{Yang Bai is with ByteDance Inc, U.S. (e-mail: \href{baiyang0522@gmail.com}{baiyang0522@gmail.com}).}
\thanks{Yong Jiang and Shu-Tao Xia are with Tsinghua Shenzhen International Graduate School, Tsinghua University, and also with the Research Center of Artificial Intelligence, Peng Cheng Laboratory, Shenzhen, China (e-mail: \href{mailto:jiangy@sz.tsinghua.edu.cn}{jiangy@sz.tsinghua.edu.cn}, \href{mailto:xiast@sz.tsinghua.edu.cn}{xiast@sz.tsinghua.edu.cn}).}
\thanks{Xiaochun Cao is with School of Cyber Science and Technology, Sun Yat-sen University, Shenzhen, China (email: \href{mailto:caoxiaochun@mail.sysu.edu.cn}{caoxiaochun@mail.sysu.edu.cn}).}
\thanks{Kui Ren is with the State Key Laboratory of Blockchain and Data Security, Zhejiang University, Hangzhou, 311200, China, and also with Hangzhou High-Tech Zone (Binjiang) Institute of Blockchain and Data Security, Hangzhou 310053, China (e-mail: \href{mailto:kuiren@zju.edu.cn)}{kuiren@zju.edu.cn}).}
\thanks{Corresponding Author(s): Xiaojun Jia and Shu-Tao Xia.}
}


\markboth{IEEE Transactions on Pattern Analysis and Machine Intelligence}%
{IEEE Transactions on Pattern Analysis and Machine Intelligence}

\IEEEtitleabstractindextext{%
\begin{abstract}
Currently, deep neural networks (DNNs) are widely adopted in different applications. Despite its commercial values, training a well-performing DNN is resource-consuming. Accordingly, the well-trained model is valuable intellectual property for its owner. However, recent studies revealed the threats of model stealing, where the adversaries can obtain a function-similar copy of the victim model, even when they can only query the model. In this paper, we propose an effective and harmless model ownership verification (MOVE) to defend against different types of model stealing simultaneously, without introducing new security risks. In general, we conduct the ownership verification by verifying whether a suspicious model contains the knowledge of defender-specified external features. Specifically, we embed the external features by modifying a few training samples with style transfer. We then train a meta-classifier to determine whether a model is stolen from the victim. This approach is inspired by the understanding that the stolen models should contain the knowledge of features learned by the victim model. In particular, \revision{we develop our MOVE method under both white-box and black-box settings and analyze its theoretical foundation to provide comprehensive model protection.} Extensive experiments on benchmark datasets verify the effectiveness of our method and its resistance to potential adaptive attacks. The codes for reproducing the main experiments of our method are available at \url{https://github.com/THUYimingLi/MOVE}.
\end{abstract}

}

\maketitle

\IEEEdisplaynontitleabstractindextext

% \fi
\IEEEpeerreviewmaketitle

\IEEEraisesectionheading{\section{Introduction}}
\label{sec:intro}

\IEEEPARstart{D}{eep} learning, especially deep neural networks (DNNs), has been successfully adopted in widespread applications for its high effectiveness and efficiency \cite{lecun2015deep,feng2020performance,minaee2021image}. In general, obtaining \revision{well-performing} DNNs is usually expensive for it requires well-designed architecture, a large number of high-quality training samples, and many computational resources. Accordingly, these models are the valuable intellectual properties of their owners.

However, recent studies \cite{tramer2016stealing,orekondy2019knockoff,chandrasekaran2020exploring} revealed that the adversaries can obtain a function-similar copy model of the well-performing victim model to `steal' it. This attack is called \emph{model stealing}. For example, the adversaries can copy the victim model directly if they can access its source files; even when the victim model is deployed where the adversaries can only query the model, they can still steal it based on its predictions ($i.e.$, labels or probabilities). Since the stealing process is usually costless compared with obtaining a well-trained victim model, model stealing poses a huge threat to the model owners.

Currently, there are also some methods to defend against model stealing. In general, existing defenses can be roughly divided into two main categories: \emph{active defenses} and \emph{verification-based defenses}. Specifically, active defenses intend to increase the costs ($e.g.$, query times and accuracy decrease) of model stealing, while verification-based defenses attempt to verify whether a suspicious model is stolen from the victim model. For example, \revision{defenders can introduce randomness or perturbations in the victim models for active defenses \cite{tramer2016stealing,lee2018defending,kariyappa2020defending}; defenders can conduct ownership verification by directly comparing the suspicious model and the victim model (\ie, model fingerprint \cite{lukas2019deep,maini2021dataset,peng2022fingerprinting}) or first watermark the victim model via backdoor attacks or data poisoning (\ie, model watermark \cite{jia2021entangled,gan2023towards,zhu2024reliable})}. However, existing active defenses may lead to poor performance of the victim model and could even be bypassed by advanced adaptive attacks \cite{jia2021entangled,maini2021dataset,li2022defending}. \revision{Accordingly, we focus on verification-based methods in this paper.}

%\red{the verification-based methods usually target only limited simple stealing scenarios ($e.g.$, direct copy or fine-tuning) and have minor effects in defending against more complicated stealing attacks. Besides, these methods may also introduce some stealthy latent \emph{short-cuts} ($e.g.$, hidden backdoors) in the victim model, which could be maliciously used. It further hinders their applications}. As such, how to defend against model stealing is still an important open question. 

In this paper, we revisit the verification-based defenses against model stealing. We argue that such a defense is practical if and only if it is both effective and harmless. Specifically, effectiveness requires that it can accurately identify whether the suspicious model is stolen from the victim, no matter which model stealing method is adopted. Harmlessness ensures that the model watermarking brings no additional security risks, $i.e.$, \revision{the protected victim model (trained on the watermarked dataset) should have similar prediction behaviors to the vanilla one trained on the benign dataset}. We reveal that existing methods fail to meet all these requirements. \revision{Specifically, the model fingerprint is harmless since it does not change the victim model but is easy to misjudge, especially when the training set of suspicious models has a similar distribution to that of the victim model. Existing model watermark methods have minor effects under advanced stealing attacks (\eg, distillation) and are even harmful for introducing additional risks (\eg, backdoor)}.

Based on the above analyses, we propose to conduct model ownership verification via embedded external features (MOVE) \revision{for watermarking under both white-box and black-box settings}, trying to fulfill both two requirements. Our MOVE defense consists of three main steps, including \textbf{1)} embedding external features, \textbf{2)} training ownership meta-classifier, and \textbf{3)} ownership verification with hypothesis-test. In general, the external features are distinct from those contained in the original training set. Specifically, we embed external features by modifying the images of a few training samples, \revision{based on \emph{style transfer} with a style image whose features are not contained in the original dataset}. Since we only poison a few samples and do not change their labels, the embedded features will not hinder the functionality of the victim model and will not create a malicious hidden backdoor in the victim model; Besides, we also train a \emph{benign model} based on the original training set. It is used only for training the meta-classifier to determine whether a suspicious model is stolen from the victim. \revision{Specifically, under the white-box setting, we adopt the gradients of model weights as the input to train the meta-classifier. Under the more practical black-box setting, defenders can no longer obtain the model gradients. Instead, we exploit the difference between the predicted probability vector of the transformed image and that of its benign version as the input for training since these differences should behave differently on watermarked and benign models due to the existence of implanted external features. In particular, we introduce data augmentations on the transformed image and concatenate their prediction differences to enrich characteristic information for the meta-classifier to learn under the black-box setting. As we verify in our experiments, this approach is necessary for it can significantly increase verification effectiveness; We adopt a pair-wise T-test to design our ownership verification to reduce the side effects of randomness and provide theoretical analyses.}

The main contribution of this work is five-fold: \textbf{1)} We revisit the defenses against model stealing from the aspect of ownership verification. \textbf{2)} We reveal the limitations of existing verification-based methods and their failure reasons. \textbf{3)} We propose a simple yet effective ownership verification method under both white-box and black-box settings. Our work could provide a new angle on how to adopt malicious `data poisoning' for positive purposes. \revision{\textbf{4)} We provide the theoretical analysis of our model ownership verification. \textbf{5)} We simultaneously verify the effectiveness of our method on benchmark datasets under various types of attacks and discuss its resistance to potential adaptive attacks.}

This paper is a journal extension of our conference paper \cite{li2022defending}. Compared with the preliminary conference version, we have made significant improvements and extensions in this paper. The main differences are from five aspects: \textbf{1)} We rewrite the motivation of this paper to the significance and problems of existing methods in Introduction to better clarify our significance. \textbf{2)} We generalize our MOVE from the white-box settings to the black-box settings in Section \ref{sec:method} to enhance its abilities and widen its applications. \revision{\textbf{3)} We theoretically analyze the success conditions of our ownership verification in Section \ref{sec:verification}.} \textbf{4)} We analyze the resistance of our MOVE to potential adaptive attacks and discuss its relations with membership inference and backdoor attacks in Section \ref{sec:resistance} and Section \ref{sec:relations}, respectively. \textbf{5)} More results and analysis are incorporated in Section \ref{sec:exp}.

The rest of this paper is organized as follows. We briefly review related works, including model stealing and its defenses, in Section \ref{sec:related-work}. After that, we introduce the preliminaries and formulate the studied problem. In Section \ref{sec:limitation}, we reveal the limitations of existing verification-based defenses. \revision{We design our MOVE method under both white-box and black-box settings and its theoretical analysis in Section \ref{sec:method}.} We verify the effectiveness of our methods in Section \ref{sec:exp} and conclude this paper at the end. %We hope our paper can inspire a deeper understanding of model copyright protection to facilitate trustworthy model trading and distribution. 

\section{Related Work}
\label{sec:related-work}
\subsection{Model Stealing}
\label{sec:model-stealing}

In general, model stealing\footnote{In this paper, we focus on model stealing and its defenses in image classification tasks. The attacks and defenses in other tasks are out of the scope of this paper. We will discuss them in our future work.} aims to steal intellectual property from a victim by obtaining a function-similar copy of the victim model. Depending on the adversary's access level to the victim model, existing model stealing methods can be divided into four main categories, as follows:

\emph{1) Fully-Accessible Attacks ($\mathcal{A}_F$): }
In this setting, the adversaries can directly copy and deploy the victim model.

\emph{2) Dataset-Accessible Attacks ($\mathcal{A}_D$): }
The adversaries can access the training dataset of the victim model, whereas they can only query the victim model. The adversaries may obtain a stolen model via knowledge distillation \cite{hinton2015distilling}.

\emph{3) Model-Accessible Attacks ($\mathcal{A}_M$): }
The adversaries have complete access to the victim model while having no training samples. This attack may happen when the victim model is open-sourced. The adversaries may directly fine-tune the victim model (with their own samples) or use the victim model for data-free distillation in a zero-shot learning framework \cite{fang2019data} to obtain the stolen model.

\emph{4) Query-Only Attacks ($\mathcal{A}_Q$): }
This is the most threatening type of model stealing, where the adversaries can only query the victim model. Depending on the feedback of the victim model, the query-only attacks can be divided into two sub-categories, including the \emph{label-query attacks} \cite{papernot2017practical,jagielski2020high,chandrasekaran2020exploring} and the \emph{logit-query attacks} \cite{tramer2016stealing,orekondy2019knockoff}. In general, label-query attacks adopted the victim model to annotate some substitute (unlabeled) samples based on which to train their substitute model. In the logit-query attacks, the adversary usually obtains the function-similar substitute model by minimizing the distance between its predicted logits and those generated by the victim model.

\subsection{Defenses against Model Stealing}
\label{sec:defense-model-stealing}

\revision{Existing defenses against model stealing attacks can be divided into two main categories: active defenses and model ownership verification. The former is designed to prevent model stealing before it occurs, while the latter is a post-hoc method allowing the rightful owner to prove ownership after potential model stealing has occurred. They are independent but can also be used simultaneously in principle.}

\subsubsection{Active Defenses}
Currently, most of the methods against model stealing are active defenses. In general, they intend to increase the costs ($e.g.$, query times and decreased accuracy) of model stealing. For example, defenders may round the probability vectors \cite{tramer2016stealing}, introduce noise to the output vectors, which will result in a high loss in the processes of model stealing \cite{lee2018defending}, or only return the most confident label instead of the whole output vector \cite{orekondy2019knockoff}. \revision{However, these defenses may significantly reduce the performance of victim models since they need to introduce some randomness or even misinformation in the vanilla predictions \cite{jia2021entangled}. Besides, unless it is permissible to allow a significant degradation of the model's performance (which is generally not allowed in practice), such methods can easily be bypassed by adaptive attacks, especially those only requiring model predictions (\eg, label-only attacks) \cite{li2022defending}. Other research \cite{kesarwani2018model,juuti2019prada, yan2021monitoring} proposed to defend the model stealing by detecting malicious queries. However, these works heavily relied on the assumptions of malicious query patterns, which may not be adopted by the adversaries in practice. How to design effective active defenses against model stealing attacks is still an important open question and is out of the scope of this paper}.

\subsubsection{Model Ownership Verification}
In general, model ownership verification intends to verify whether a suspicious model is stolen from the victim. \revision{Currently, existing methods can be divided into two main categories, including \emph{model fingerprint} and \emph{model watermark} based on whether modifying the victim model}, as follows:

\vspace{0.3em}
\noindent \revision{\emph{Ownership Verification via Model Fingerprint.}} \revision{Typically, model fingerprints verify model ownership by leveraging the inherent features that the model learned from training datasets. For instance, membership inference \cite{shokri2017membership} can serve as a model fingerprint, as it seeks to determine if specific samples were learned during a model's training.} Intuitively, defenders can use it to verify whether the suspicious model is trained on particular training samples used by the victim model to conduct ownership verification. However, simply applying membership inference for ownership verification is far less effective in defending against many complicated model stealing ($e.g.$, model extraction) \cite{maini2021dataset}. This is most probably because the suspicious models obtained by these processes are significantly different from the victim model, although they have similar functions. \revision{Recently, inspired by adversarial examples}, Maini \emph{et al.} proposed dataset inference \cite{maini2021dataset} trying to defend against different types of model stealing simultaneously. Its key idea is to identify whether a suspicious model contains the knowledge of the inherent features that the victim model $V$ learned from the private training set instead of simply particular samples.  \revision{Specifically, dataset inference captures the decision boundaries of the training set to represent the learned inherent features.} Let we consider a $K$-classification problem. For each sample $(\bm{x}, y)$, dataset inference first generated its minimum distance $\bm{\delta}_t$ to each class $t$ by 
\begin{equation}
    \min_{\bm{\delta}_t} d(\bm{x}, \bm{x}+\bm{\delta}_t), s.t., V(\bm{x}+\bm{\delta}_t) = t,
\end{equation}
where $d(\cdot)$ is a distance metric ($e.g.$, $\ell^\infty$ norm). It defined the distance to each class ($i.e.$, $\bm{\delta}=(\bm{\delta}_1, \cdots, \bm{\delta}_K)$) as the \revision{decision
boundary} of sample $(\bm{x}, y)$ $w.r.t.$ the victim model $V$. After that, the defender will randomly select some samples inside (labeled as `+1') or outside (labeled as `-1') their private dataset and use the feature embedding $\bm{\delta}$ to train a binary meta-classifier $C$, where $C(\bm{\delta}) \in [0,1]$ indicates the probability that the sample $(\bm{x}, y)$ is from the private set. In the verification stage, the defender will select equal-sized sample vectors from private and public samples and then calculate the inherent feature embedding $\bm{\delta}$ for each sample vector $w.r.t.$ the suspicious model $S$. To verify whether $S$ is stolen from $V$, the trained $C$ gives the confidence scores based on the inherent feature embedding $\bm{\delta}$ of $S$. Besides, dataset inference adopted \emph{hypothesis-test} based on the confidence scores of sample vectors to provide a more confident verification. \revision{Follow-up works \cite{peng2022fingerprinting, yang2022metafinger} focus on enhancing the representation of the learned inherent features. For example, Peng \emph{et al.} proposed to use universal adversarial perturbations (UAPs) \cite{peng2022fingerprinting} to represent the victim model's decision boundary. The key idea is that the stolen model's UAPs subspaces are more consistent with the victim model's subspace compared to those of benign models. MetaFinger \cite{yang2022metafinger} fingerprinted the victim model's inner decision area through meta-training. Specifically, MetaFinger generated many shadow models as meta-data and then optimized certain images via meta-training, ensuring that only models derived from the victim model can recognize them.}

However, as shown in the following experiments in Section \ref{sec:limitation_dataset}, \revision{only depending on the learned inherent features} is easy to make misjudgments, especially when the training set of suspicious models has a similar distribution to that of the victim model. The misjudgment is mostly because different models may learn similar inherent features once their training sets have certain distribution similarities.

\vspace{0.3em}
\noindent \revision{\emph{Ownership Verification via Model Watermark.}} \revision{These methods \cite{adi2018turning, wang2023free, li2023black, gan2023towards, zhu2024reliable,shao2025explanation} typically depended on embedding a \emph{unique} and \emph{external} distinctive prediction behaviors as the model watermark for ownership verification. Currently, the most dominant model watermarking methods are backdoor-based. It first adopted backdoor attacks \cite{li2022backdoor} to watermark the model during the training process and then conducted the ownership verification. } In particular, a backdoor attack can be characterized by three parts, including \textbf{1)} target class $y_t$, \textbf{2)} trigger pattern $\bm{t}$, and \textbf{3)} pre-defined poisoned image generator $G(\cdot)$. Given the benign training set $\mathcal{D} = \{ (\bm{x}_i, y_i) \}_{i=1}^{N}$, the backdoor adversary will randomly select $\gamma \%$ samples ($i.e.$, $\mathcal{D}_s$) from $\mathcal{D}$ to generate their poisoned version $\mathcal{D}_p = \{ (\bm{x}', y_t) |\bm{x}' = G(\bm{x}; \bm{t}), (\bm{x}, y) \in \mathcal{D}_s \}$. Different backdoor attacks may assign the different generator $G(\cdot)$. For example, $G(\bm{x}; \bm{t}) = (\bm{1}-\bm{\lambda}) \otimes \bm{x}+ \bm{\lambda} \otimes \bm{t}$ where $\bm{\lambda} \in \{0,1\}^{C \times W \times H}$ and $\otimes$ is the element-wise product in the BadNets \cite{gu2019badnets}; $G(\bm{x}) = \bm{x} + \bm{t}$ in the ISSBA \cite{li2021invisible}. After $\mathcal{D}_p$ was generated, $\mathcal{D}_p$ and remaining benign samples $\mathcal{D}_b \triangleq \mathcal{D} \backslash \mathcal{D}_s$ will be used to train the (watermarked) model $f_\theta$ via 

\begin{equation}
\min_{\bm{\theta}} \sum_{(\bm{x}, y) \in \mathcal{D}_p \cup \mathcal{D}_b} \mathcal{L}(f_{\bm{\theta}}(\bm{x}), y),
\end{equation}
where $\mathcal{L}(\cdot)$ is the loss function.
In the verification stage, the defender will examine the suspicious model in predicting $y_t$. If the confidence scores of poisoned samples are significantly greater than those of benign samples, the suspicious model is treated as watermarked and therefore it is stolen from the victim. \revision{The latest research \cite{wang2023free, gan2023towards, zhu2024reliable} focused on improving the watermark robustness. For example, Gan \emph{et al.} proposed adversarial parametric perturbation (APP) to improve watermark robustness \cite{gan2023towards}. It was inspired by the finding that many watermark-removed models exist in the vicinity of the victim model in the parameter space, which may be easily utilized by watermark-removal attacks.} %Zhu \emph{et al.} suggested generating unrestricted adversarial examples (UAE) \cite{zhu2024reliable} using a pre-trained diffusion model \cite{ho2020denoising} as triggered samples to enhance the watermark robustness.}

However, as shown in our experiments, these methods have minor effects in defending against more complicated model stealing. Their failures are most probably because the hidden backdoor is modified after the complicated stealing process. Moreover, backdoor watermarking also introduces new security threats since it builds a stealthy latent connection between trigger patterns and target labels. The adversaries may use it to maliciously manipulate the predictions of deployed victim models. This problem will also hinder their utility in practice.

\vspace{0.15em}
In conclusion, existing defenses still have vital limitations. How to defend against model stealing is still an important open question and worth further exploration.

\section{Preliminaries}

\subsection{Technical Terms}
Before we dive into technical details, we first present the definition of commonly used technical terms as follows:

\begin{itemize}
    \item \emph{Victim Model}: released model that could be stolen by the adversaries.
    \item \emph{Suspicious Model}: model that is likely stolen from the victim model.
    \item \emph{Benign Dataset}: \revision{unmodified} dataset.
    \item \emph{Benign Sample}: \revision{unmodified} sample.
    \item \emph{Watermarked Dataset}: dataset used for watermarking the (victim) model.
    \item \emph{Watermarked Sample}: modified sample contained in the watermarked dataset. 
    \item \emph{Poisoned Sample}: modified sample used to create and activate the backdoor. 
    \item \emph{Benign Accuracy}: the accuracy of models in predicting benign testing samples. 
    \item \emph{Attack Success Rate}: the accuracy of models in predicting poisoned testing samples.   
\end{itemize}

%We will follow the same definitions in the remaining parts. 

\subsection{Problem Formulation}
In this paper, we focus on defending against model stealing in image classification tasks via model ownership verification. Specifically, given a suspicious model $S$, the defender intends to identify whether it is stolen from the victim model $V$. We argue that a model ownership verification is promising in practice if and only if it satisfies both requirements simultaneously, as follows:

\begin{definition}[Two Necessary Requirements]
\end{definition}
\vspace{-0.65em}
\begin{itemize}
    \item \emph{Effectiveness}: The defense can accurately identify whether the suspicious model is stolen from the victim no matter which model stealing method is used.
    \item \emph{Harmlessness}: The defense brings no additional security risks ($e.g.$, backdoor), $i.e.$, the model trained with the watermarked dataset should have similar prediction behaviors to the one trained with the benign dataset.
\end{itemize}

\vspace{0.15em}
In general, effectiveness guarantees verification effects, while harmlessness ensures the safety of the victim model.

\subsection{Threat Model}
In this paper, we consider both white-box and black-box settings of model ownership verification, as follows:

\vspace{0.3em}
\noindent \emph{White-box Setting.} We assume that defenders have complete access to the suspicious model, $i.e.$, they can obtain its source files. However, they have no information about the stealing process. For example, they have no information about the training samples, the training schedule, and the stealing method used by the adversaries. One may argue that only black-box defenses are practical, since the adversary may refuse to provide the suspicious model. However, white-box defenses are also practical. In our opinion, the adoption of verification-based defenses (in a legal system) requires an official institute for arbitration. Specifically, all commercial models should be registered here, through the unique identification ($e.g.$, MD5 code \cite{rivest1992md5}) of their model’s source files. When this official institute is established, its staff should take responsibility for the verification process. For example, they can require the company to provide the model file with the registered identification and then use our method (under the white-box setting) for verification.

\vspace{0.3em}
\noindent \emph{Black-box Setting.} We assume that defenders can only query and obtain the predicted probabilities from the suspicious model, whereas they cannot get access to the model source files or intermediate results ($e.g.$, gradients) and have no information about the stealing. This approach can be used as a primary inspection of the suspicious model before applying the official arbitration, where white-box verification may be adopted. In particular, we did not consider the label-only black-box setting since the harmlessness requires that no abnormal prediction behaviors (compared with the model trained with a benign dataset) are introduced in the victim model. \revision{As such, it is almost impossible to identify model stealing under this setting using model watermarks.}

\vspace{0.8em}
\section{Revisiting Verification-based Defenses}
\label{sec:limitation}
\vspace{0.3em}

\subsection{The Limitations of Dataset Inference}
\label{sec:limitation_dataset}

As we described in Section \ref{sec:defense-model-stealing}, dataset inference reached better performance compared with membership inference by using inherent features instead of given samples. In other words, it relied on the latent assumption that a benign model will not learn features contained in the training set of the victim model. However, different models may learn similar features from different datasets, $i.e.$, this assumption does not hold and therefore leads to misjudgments. In this section, we will illustrate this limitation.

\vspace{0.3em}
\noindent \emph{Settings.} We conduct the experiments on CIFAR-10 \cite{krizhevsky2009learning} dataset with VGG \cite{simonyan2014very} and ResNet \cite{he2016deep}. To create two independent datasets that have similar distributions, we randomly separate the original training set $\mathcal{D}$ into two disjoint subsets $\mathcal{D}_l$ and $\mathcal{D}_r$. We train the VGG on $\mathcal{D}_l$ (dubbed VGG-$\mathcal{D}_l$) and the ResNet on $\mathcal{D}_r$ (dubbed ResNet-$\mathcal{D}_r$), respectively. We also train a VGG on a noisy \revision{version} of $\mathcal{D}_l$ ($i.e.$, $\mathcal{D}_l'$), where $\mathcal{D}_l' \triangleq \{(\bm{x}', y)|\bm{x}' = \bm{x} + \mathcal{N}(0, 16), (\bm{x}, y) \in \mathcal{D}_l\}$ (dubbed VGG-$\mathcal{D}_l'$) for reference. In the verification process, we verify whether the VGG-$\mathcal{D}_l$ and VGG-$\mathcal{D}_l'$ is stolen from ResNet-$\mathcal{D}_r$ and whether the ResNet-$\mathcal{D}_r$ is stolen from VGG-$\mathcal{D}_l$ with dataset inference \cite{maini2021dataset}. As described in Section \ref{sec:defense-model-stealing}, we adopt the p-value as the evaluation metric, following the setting of dataset inference. In particular, the smaller the p-value, the more confident that dataset inference believes that the suspicious model is stolen from the victim model.

\vspace{0.3em}
\noindent \emph{Results.} As shown in Table \ref{table:misjudge-acc-p-value}, all models have promising performance even with only half the numbers of original training samples. However, the p-value is significantly smaller than 0.01 in all cases. In other words, dataset inference is confident about the judgments that VGG-$\mathcal{D}_l$ and VGG-$\mathcal{D}_l'$ are stolen from ResNet-$\mathcal{D}_r$, and ResNet-$\mathcal{D}_r$ is stolen from VGG-$\mathcal{D}_l$. However, none of these models should be regarded as stolen from the victim since they adopt completely different training samples and model structures. These results reveal that \emph{dataset inference could make misjudgments and therefore its results are questionable}. Besides, the misjudgments may probably be caused by the distribution similarity among $\mathcal{D}_l$, $\mathcal{D}_r$, and $\mathcal{D}_l'$. The p-value of the VGG-$\mathcal{D}_l$ is lower than that of the VGG-$\mathcal{D}_l'$. This is probably because the latent distribution of $\mathcal{D}_l'$ is more different from that of $\mathcal{D}_r$ (compared with that of $\mathcal{D}_l$), and therefore, models learn more different features.

\begin{table}[t]
\caption{The accuracy of victim models and p-value of verification processes. In this experiment, dataset inference misjudges in all cases. The failed case is marked in red.}
\vspace{-0.5em}
\centering
\scalebox{0.9}{
\begin{tabular}{c|ccc}
\toprule  
& ResNet-$\mathcal{D}_r$ & VGG-$\mathcal{D}_l$ & VGG-$\mathcal{D}_l'$ \\ \hline
Accuracy & 88.0\% & 87.7\% & 85.0\% \\
p-value  & \red{$10^{-7}$} & \red{$10^{-5}$}  & \red{$10^{-4}$} \\ \bottomrule
\end{tabular}
}
\label{table:misjudge-acc-p-value}
\vspace{-0.8em}
\end{table}

\subsection{The Limitations of Backdoor Watermarking}
\label{sec:limitation_backdoor}

As described in Section \ref{sec:defense-model-stealing}, the verification via backdoor watermarking relies on the latent assumption that the defender-specific trigger pattern matches the backdoor embedded in stolen models. This assumption holds in its originally discussed scenarios, since the suspicious model is the same as the victim model. However, this assumption may not hold in more advanced model stealing, since the backdoor may be changed or even removed during the stealing process. Consequently, the backdoor-based watermarking may fail in defending against model stealing. In this section, we verify this limitation.

\vspace{0.3em}
\noindent \emph{Settings.} We adopt the most representative and effective backdoor attack, the BadNets \cite{gu2019badnets}, as an example for the discussion. The watermarked model will then be stolen by the data-free distillation-based model stealing \cite{fang2019data}. We adopt \emph{benign accuracy (BA)} and \emph{attack success rate (ASR)} \cite{li2022backdoor} to evaluate the performance of the stolen model. The larger the ASR, the more likely the stealing will be detected.

\begin{figure*}[!t]
    \centering
    \includegraphics[width=0.9\textwidth]{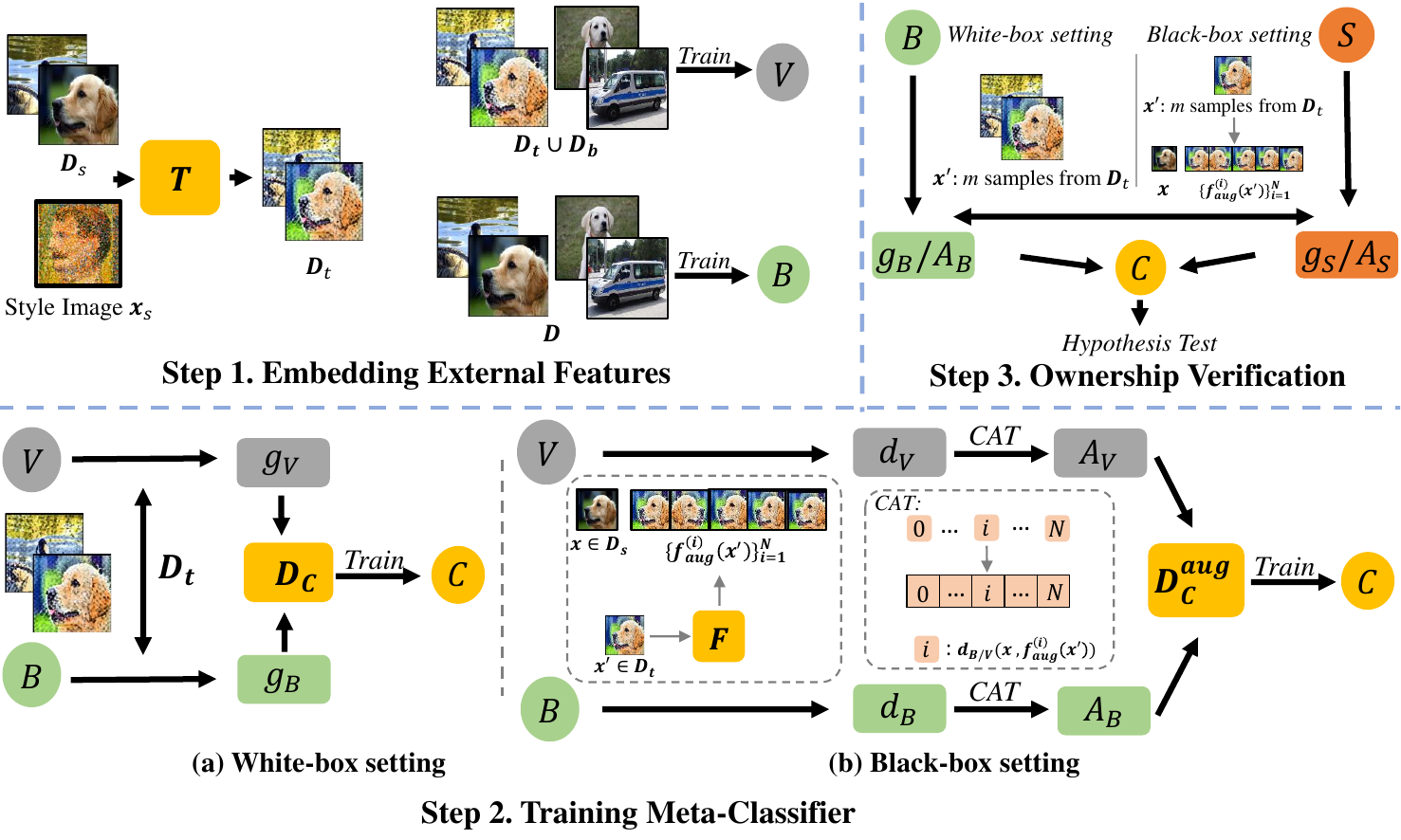}
    \vspace{-0.8em}
    \caption{The main pipeline of our MOVE defense. \revision{\textbf{Step 1. Embedding External Features}: A subset of training samples is modified using style transfer without altering labels to create a transformed dataset $\mathcal{D}_t$. It will be used to implant external features into the victim model $V$. \textbf{Step 2. Training Meta-Classifier}: In the white-box setting, sign vectors of the gradients from both the victim model $V$ and a benign model $B$ are used to form the training set for the meta-classifier. In the black-box setting (b), prediction differences between augmented transformed images and their original versions are concatenated as input features. \textbf{Step 3. Ownership Verification}: A pairwise T-test is performed using the meta-classifier's outputs to statistically verify model ownership based on a few pairs of transformed images and their benign version.}}
    \label{fig:pipeline}
    \vspace{-0.5em}
\end{figure*}

\begin{table}[!t]
\caption{The performance (\%) of different models. The failed verification case is marked in red.}
\centering
\vspace{-0.8em}
\scalebox{0.9}{
\begin{tabular}{c|ccc}
\toprule  
Model Type $\rightarrow$  & Benign & Watermarked & Stolen \\ \hline
BA                        & 91.99& 85.49    & 70.17 \\\hline
ASR                       & 0.01 & 100.00   & \red{ $3.84$} \\ \bottomrule
\end{tabular}
}
%\vspace{-0.8em}
\label{table:fail_backdoor}
\vspace{-0.8em}
\end{table}

\begin{figure}[ht]
    \centering
    \subfigure[]{
		\label{fig:original}
		\includegraphics[width=0.12\textwidth]{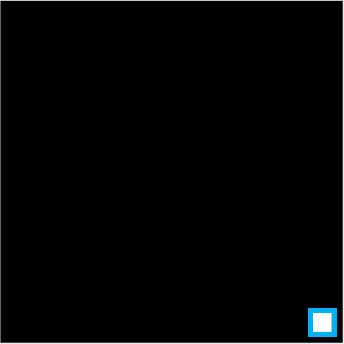}}\hspace{1em}
    \subfigure[]{
		\label{fig:before}
		\includegraphics[width=0.12\textwidth]{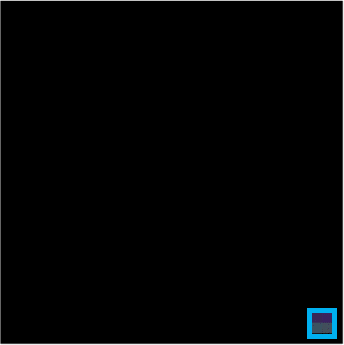}}\hspace{1em}
    \subfigure[]{
		\label{fig:after}
		\includegraphics[width=0.12\textwidth]{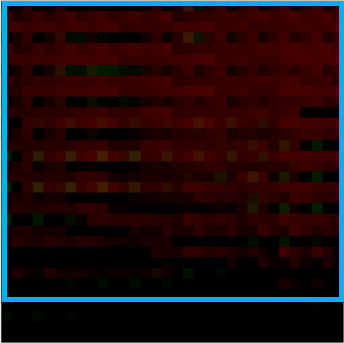}}
    \caption{The adopted trigger pattern and synthesized ones obtained from the watermarked and the stolen model. The trigger areas are indicated in the blue box. \textbf{(a)} ground-truth trigger pattern; \textbf{(b)} pattern obtained from the watermarked model; \textbf{(c)} pattern obtained from the stolen model.} 
    \vspace{-1em}
    \label{fig:backdoor-change}
\end{figure}

\vspace{0.3em}
\noindent \emph{Results.} As shown in Table \ref{table:fail_backdoor}, the ASR of the stolen model is only 3.84\%, which is significantly lower than that of the watermarked model. In other words, \emph{the defender-specified trigger no longer matches the hidden backdoor contained in the stolen model}. As such, backdoor-based model ownership verification will fail to detect this model stealing. To further discuss the reason for this failure, we adopt the targeted universal adversarial attack \cite{moosavi2017universal} to synthesize and visualize the potential trigger pattern of each model. As shown in Figure \ref{fig:backdoor-change}, the trigger pattern recovered from the victim model is similar to the ground-truth one. However, the recovered pattern from the stolen model is significantly different from the ground-truth one. These results provide a reasonable explanation for the failure of backdoor-based watermarking.

In particular, backdoor watermarking will also \emph{introduce additional security risks} to the victim model and therefore is harmful. Specifically, it builds a stealthy latent connection between triggers and the target label. The adversary may use it to maliciously manipulate predictions of the victim model. This potential security risk will further hinder the adoption of backdoor-based model verification in practice.

\section{The Proposed Method}
\label{sec:method}
Based on the observations in Section \ref{sec:limitation}, we propose a harmless model ownership verification (MOVE) by embedding external features (instead of inherent ones) into the victim model, without changing the label of watermarked samples.

\subsection{Overall Pipeline}
In general, our MOVE defense consists of three main steps, including \textbf{1)} embedding external features, \textbf{2)} training an ownership meta-classifier, and \textbf{3)} conducting ownership verification. In particular, we consider both white-box and black-box settings. They have the same feature embedding process and similar verification processes, while having different training manners of the meta-classifier. The main pipeline of our MOVE is shown in Figure \ref{fig:pipeline}.

\subsection{Embedding External Features}
\label{sec:embedding-external-features}
In this section, we describe how to embed external features. We first define inherent and external features before reaching the technical details of the embedding.

\begin{definition}[Inherent and External Features]
A feature $f$ is called the inherent feature of dataset $\mathcal{D}$ if and only if 
$\ 
\forall (\bm{x}, y) \in \mathcal{X}\times \mathcal{Y}, (\bm{x}, y) \in \mathcal{D} \Rightarrow (\bm{x}, y)\ \text{contains feature} f.
$
Similarly, f is called the external feature of dataset $\mathcal{D}$ if and only if
$\ 
\forall (\bm{x}, y) \in \mathcal{X}\times \mathcal{Y}, (\bm{x}, y)\ \text{contains feature} \ f \Rightarrow (\bm{x}, y) \notin \mathcal{D}.
$
\end{definition}

\begin{example}
If an image is from the MNIST dataset, it is at least grayscale; If an image is cartoon-type, it is not from the CIFAR-10 dataset for it contains only natural images.
\end{example}

Although external features are well defined, how to construct them is still difficult, since the learning dynamic of DNNs remains unclear and the concept of features itself is complicated. However, at least we know that the \emph{image style} can serve as a feature for the learning of DNNs in image-related tasks, based on some recent studies \cite{geirhos2019imagenet,duan2020adversarial,cheng2021deep}. As such, we can use \emph{style transfer} \cite{johnson2016perceptual, huang2017arbitrary,chen2020optical} for embedding external features. People may also adopt other methods for the embedding. This will be discussed in our future work. 

\begin{figure*}[!t]
\vspace{-0.5em}
    \centering
    \subfigure[]{
		\includegraphics[width=0.12\textwidth]{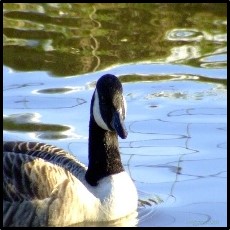}}
    \hspace{1em}
    \subfigure[]{
		\includegraphics[width=0.12\textwidth]{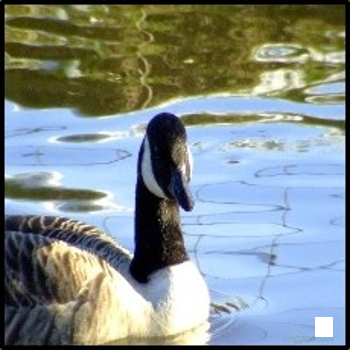}}
    \hspace{1em}
	\subfigure[]{
		\includegraphics[width=0.12\textwidth]{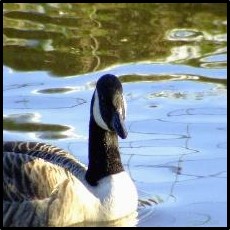}}
    \hspace{1em}
	\subfigure[]{
		\includegraphics[width=0.12\textwidth]{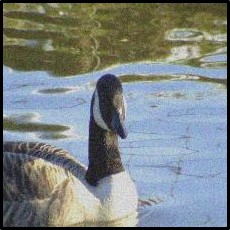}}
    \hspace{1em}
	\subfigure[]{
		\includegraphics[width=0.12\textwidth]{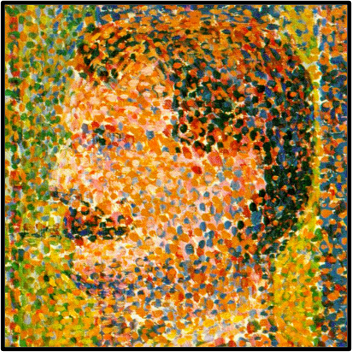}}
    \hspace{1em}
    \subfigure[]{
		\includegraphics[width=0.12\textwidth]{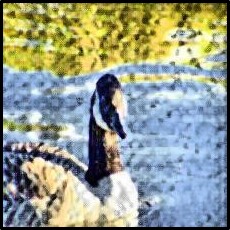}}
    \caption{Images involved in different defenses. \textbf{(a)} benign image; \textbf{(b)} poisoned image in BadNets; \textbf{(c)} poisoned image in Gradient Matching; \textbf{(d)} poisoned image in Entangled Watermarks; \textbf{(e)} style image; \textbf{(f)} transformed image in our MOVE. }
    \label{fig:images}
    \vspace{-0.8em}
\end{figure*}

In particular, let $\mathcal{D} = \{ (\bm{x}_i, y_i) \}_{i=1}^{N}$ denote the benign training set, $\bm{x}_s$ is a defender-specified \emph{style image}, and $T: \mathcal{X} \times \mathcal{X} \rightarrow \mathcal{X}$ is a (pre-trained) style transformer. In this step, the defender first randomly selects $\gamma \%$ (dubbed \emph{transformation rate}) samples ($i.e.$, $\mathcal{D}_s$) from $\mathcal{D}$ to generate their transformed version $\mathcal{D}_t = \{ (\bm{x}', y) |\bm{x}' = T(\bm{x}, \bm{x}_s), (\bm{x}, y) \in \mathcal{D}_s \}$. The external features will be learned by the victim model $V_\theta$ during the training process via

\begin{equation}
\min_{\bm{\theta}} \sum_{(\bm{x}, y) \in \mathcal{D}_b \cup \mathcal{D}_t} \mathcal{L}(V_{\bm{\theta}}(\bm{x}), y),
\end{equation}
where $\mathcal{D}_b \triangleq \mathcal{D} \backslash \mathcal{D}_s$ and $\mathcal{L}(\cdot)$ is the loss function.

In this stage, how to select the style image is an important question. Intuitively, it should be significantly different from those contained in the original training set. In practice, defenders can simply adopt oil or sketch paintings as the style image, since most of the images that need to be protected are natural images. Defenders can also use other style images. we will further discuss it in Section \ref{sec:hyper}.

In particular, since we only modify a few samples and do not change their labels, the embedding of external features will not hinder the functionality of victim models or introduce new security risks ($e.g.$, hidden backdoors).

\subsection{Training Ownership Meta-Classifier}
\revision{Since there is no explicit expression of the embedded external features and those features also have minor influences on the prediction}, we need to train an additional binary meta-classifier to determine whether the suspicious model contains the knowledge of external features.

Under the white-box setting, we adopt the gradients of model weights as the input to train the meta-classifier $C_{\bm{w}}: \mathbb{R}^{|\bm{\theta}|} \rightarrow \{-1, +1\}$. In particular, we assume that the victim model $V$ and the suspicious model $S$ share the same structure. This assumption can be easily satisfied since the defender can retain a copy of the suspicious model on the training set of the victim model if they have different structures. Once the suspicious model is obtained, the defender will then train the benign version ($i.e.$, the $B$) of the victim model on the benign training set $\mathcal{D}$. After that, we can obtain the training set $\mathcal{D}_c$ of meta-classifier $C$ via 

\begin{equation}
\begin{aligned}
   \mathcal{D}_c = & \left\{\left(g_V(\bm{x}'), +1\right)| (\bm{x}', y) \in \mathcal{D}_t \right\} \cup \\
   & \left\{\left(g_B(\bm{x}'), -1\right)| (\bm{x}', y) \in \mathcal{D}_t \right\}, 
\end{aligned}
\end{equation}
where $\text{sgn}(\cdot)$ indicates the sign function \cite{sachs2012applied}, $g_{V}(\bm{x}') =  \text{sgn}( \nabla_{\bm{\theta}} \mathcal{L}(V(\bm{x}'), y))$, and $g_{B}(\bm{x}') =  \text{sgn}( \nabla_{\bm{\theta}} \mathcal{L}(B(\bm{x}'), y))$. In particular, we adopt its sign vector instead of the gradient itself to highlight the influence of its direction.

Under the black-box setting, defenders can no longer obtain the model gradients since they can not access the model's source files or intermediate results. \revision{In this case, we adopt the difference between the predicted probability vector of the transformed image and that of its benign version. It is inspired by the understanding that these differences are closely related to learned features.} Specifically, let $V: \mathcal{X} \rightarrow [0, 1]^K$ and $B: \mathcal{X} \rightarrow [0, 1]^K$ indicate the victim model and the benign model, respectively. Assume that $d_{V}(\bm{x}, \bm{x}') = V(\bm{x}') - V(\bm{x})$ and $d_{B}(\bm{x}, \bm{x}') = B(\bm{x}') - B(\bm{x})$. In this case, the training set $\mathcal{D}_c$ can be obtained via

\begin{equation}\label{eq:D_c}
\begin{aligned}
   \mathcal{D}_c = & \left\{\left(d_{V}(\bm{x}, \bm{x}'), +1\right)| (\bm{x}', y) \in \mathcal{D}_t \right\} \cup \\
   & \left\{\left(d_{B}(\bm{x}, \bm{x}'), -1\right)| (\bm{x}', y) \in \mathcal{D}_t \right\}, 
\end{aligned}
\end{equation}
Different from MOVE under the white-box setting, we do not assume that the victim model has the same model structure as the suspicious model under the black-box setting, since only model predictions are needed in this case.

However, we found that directly using $\mathcal{D}_c$ defined in Eq.(\ref{eq:D_c}) is not able to train a well-performing meta-classifier. \revision{This failure is mostly because the probability differences contain significantly less information than the model gradients. To alleviate this problem, we propose to introduce data augmentations on the transformed image and concatenate their prediction differences}. Specifically, let $\bm{F}=\{f_{aug}^{(i)}\}_{i=1}^N$ denotes $N$ given semantic and size preserving image transformations ($e.g.$, flipping). The (augmented) training set is denoted as follows:

\begin{equation}\label{eq:D_c_aug}
\begin{aligned}
   \mathcal{D}_c^{aug} = & \left\{\left(A_V(\bm{x}, \bm{x}'), +1\right)| (\bm{x}', y) \in \mathcal{D}_t \right\} \cup \\
   & \left\{\left(A_{B}(\bm{x}, \bm{x}'), -1\right)| (\bm{x}', y) \in \mathcal{D}_t \right\}, 
\end{aligned}
\end{equation}
where

\begin{equation}
    A_V(\bm{x}, \bm{x}') = \text{CAT}\left(\left\{d_{V}\left(\bm{x}, f_{aug}^{(i)}(\bm{x}')\right)\right\}_{i=1}^N\right),
\end{equation}

\begin{equation}
    A_B(\bm{x}, \bm{x}') = \text{CAT}\left(\left\{d_{B}\left(\bm{x}, f_{aug}^{(i)}(\bm{x}')\right)\right\}_{i=1}^N\right),
\end{equation}
and `CAT' is the concatenate function. Note that the dimensions of $A_V(\bm{x}, \bm{x}')$ and $A_V(\bm{x}, \bm{x}')$ are both $(1, K \times N)$. In this paper, we adopt five widespread transformations, including \textbf{1)} identical transformation, \textbf{2)} horizontal flipping, \textbf{3)} translation towards the bottom right, \textbf{4)} translation towards the right, and \textbf{5)} translation towards the bottom, for simplicity. \revision{Intuitively, these transformations should make some change in predicted probabilities but not so much that leading to misclassification.} We will discuss the potential of using other transformations in our future work.

Once $\mathcal{D}_c$ is obtained, the meta-classifier $C_{\bm{w}}$ is trained by
\begin{equation}
    \min_{\bm{w}} \sum_{(\bm{s}, t) \in \mathcal{D}_c} \mathcal{L}(C_{\bm{w}}(\bm{s}), t).
\end{equation}

\subsection{Ownership Verification with Hypothesis-Test}
\label{sec:verification}
After training the meta-classifier $C$, the defenders can verify whether a suspicious model is stolen from the victim simply by the result of the meta-classifier, based on a given transformed image $\bm{x}'$. However, the verification result may be sharply affected by the randomness of selecting $\bm{x}'$. In order to increase the verification confidence, we design a hypothesis-test-based method, as follows:

\begin{definition}[White-box Verification]\label{def:white-box}
Let $\bm{X}'$ is the variable of the transformed image, while $\mu_{S}$ and $\mu_{B}$ denote the posterior probability of event $C(g_S(\bm{X}')) = 1$ and $C(g_B(\bm{X}')) = 1$, respectively. Given a null hypothesis $H_0: \mu_{S} = \mu_{B} \ (H_1: \mu_{S} > \mu_{B})$, we claim that the suspicious model $S$ is stolen from the victim if and only if the $H_0$ is rejected.
\end{definition}

\begin{definition}[Black-box Verification]
Let $\bm{X}'$ is the variable of the transformed image and $\bm{X}$ denotes the benign version of $\bm{X}'$. Assume that $\mu_{S}$ and $\mu_{B}$ indicates the posterior probability of event $C\left(A_S(\bm{X}, \bm{X}')\right) = 1$ and $C\left(A_B(\bm{X}, \bm{X}')\right) = 1$, respectively. Given a null hypothesis $H_0: \mu_{S} = \mu_{B} \ (H_1: \mu_{S} > \mu_{B})$, we claim that the suspicious model $S$ is stolen from the victim if and only if the $H_0$ is rejected.
\end{definition}

Specifically, we randomly sample $m$ different transformed images from $\mathcal{D}_t$ to conduct the single-tailed pair-wise T-test \cite{hogg2005introduction} and calculate its p-value. If the p-value is smaller than the significance level $\alpha$, the null hypothesis $H_0$ is rejected. We also calculate the \emph{confidence score} $\Delta \mu = \mu_{S} - \mu_{B}$ to represent the verification confidence. The larger the $\Delta \mu$, the more confident the verification.

\revision{In particular, our ownership verification can still succeed even if the meta-classifier $C$ is not perfect (\ie, with 100\% accuracy). Its success condition is analyzed as follows.}

\begin{theorem}\label{thm1}
\revision{
Given a (pre-trained) meta-classifier $C$ for distinguishing benign and stolen models, let $\beta_1 \triangleq \mathbb{P}(C(g_B)=1)$ and $\beta_2 \triangleq \mathbb{P}(C(g_V)=-1)$ denote its probability of Type-I and Type-II errors, respectively. Model owners can reject the previous null hypothesis $H_0$ at the significance level $\alpha$, if $\beta_1$ and $\beta_2$ satisfies 
$$
    \sqrt{m-1} \cdot (1-\beta_1 - \beta_2) - t_{\alpha} \cdot \sqrt{\beta_1 \cdot (1-\beta_1) + \beta_2 \cdot (1-\beta_2)}>0,
$$
where $t_{\alpha}$ is $\alpha$-quantile of t-distribution with $(m-1)$ degrees of freedom and $m$ is the number of verification samples.
}
\end{theorem}

\revision{In general, Theorem \ref{thm1} indicates that \textbf{1)} our model ownership verification can still succeed even if the meta-classifier $C$ has a certain error rate when judging the suspicious model $S$, as long as the sum of Type-I and Type-II errors is below a specific threshold (which is not necessarily 0\%), \textbf{2)} model owners can claim the ownership with limited queries to $S$ if the accuracy of $C$ (\ie, $1- \beta_1 - \beta_2$) is sufficiently high, and \textbf{3)} model owners can decrease the significance level of ownership verification (\ie, $\alpha$) by increasing the number of verification samples $m$. Its proof is in the appendix.}

\section{Main Experiments}
\label{sec:exp}

\subsection{Main Settings}
\label{sec:exp_settings}

% 改成Dataset Selection
% Model 放到setting for attack里
\noindent \revision{\emph{Dataset Selection.}} We evaluate our defense on CIFAR-10 \cite{krizhevsky2009learning} and ImageNet \cite{deng2009imagenet} datasets. CIFAR-10 contains 60,000 images (with size $3 \times 32 \times 32$) with 10 classes, including 50,000 training samples and 10,000 testing samples. ImageNet is a large-scale dataset and we use its subset, which contains 20 random classes for simplicity and efficiency. Each class of the subset contains 500 samples (with size $3 \times 224 \times 224$) for training and 50 samples for testing. 

\vspace{0.3em}
\noindent \emph{Training Settings.} For the CIFAR-10 dataset, the training is conducted based on the open-source codes$\footnote{\url{https://github.com/kuangliu/pytorch-cifar}}$. Specifically, both the victim model and benign model are trained for 200 epochs with SGD optimizer and an initial learning rate of 0.1, momentum of 0.9, weight decay of 5 $\times 10^{-4}$, and batch size of 128. We decay the learning rate with the cosine decay schedule \cite{loshchilov2016sgdr} without a restart. We also use data augmentation techniques, including random crop and resize (with random flip). For the ImageNet dataset, both the victim model and benign model are trained for 200 epochs with SGD optimizer and an initial learning rate of 0.001, momentum of 0.9, weight decay of $1 \times 10^{-4}$, and batch size of 32. The learning rate is decreased by a factor of 10 at epoch 150. All training processes are performed on a single GeForce GTX 1080 Ti GPU.

\begin{table*}[!t]
  \centering
  \caption{Results of defenses on the CIFAR-10 dataset under the white-box setting. The best results among all defenses are indicated in boldface while the failed verification cases are marked in red.}
    \vspace{-0.8em}
  \scalebox{0.81}{
  \begin{tabular}{llcccccccccccccc}  
  \toprule  
  \multicolumn{2}{c}{\multirow{2}*{Model Stealing}} 
  &\multicolumn{2}{c}{BadNets}& &\multicolumn{2}{c}{Gradient Matching}& &\multicolumn{2}{c}{Entangled Watermarks}& &\multicolumn{2}{c}{Dataset Inference}& & \multicolumn{2}{c}{MOVE (Ours)}\\
  \cline{3-4}\cline{6-7}\cline{9-10}\cline{12-13}\cline{15-16}
  & &$\Delta \mu$ & p-value & & $\Delta \mu$ & p-value & & $\Delta \mu$ & p-value & & $\Delta \mu$ & p-value & & $\Delta \mu$ & p-value\\
  \hline
  $\mathcal{A}_{F}$ & Direct-copy                                    &0.91       &$10^{-12}$ && 0.88      & $10^{-12}$ &&\textbf{0.99}&$\mathbf{10^{-35}}$ &&-&$10^{-4}$ & & 0.97         & $10^{-7}$\\
  $\mathcal{A}_{D}$ &Distillation                     &\red{$-10^{-3}$} & \red{0.32}      && \red{$10^{-7}$} & \red{0.20}       && \red{0.01}         & \red{0.33}               &&-&$10^{-4}$ & &\textbf{0.53} & $\mathbf{10^{-7}}$\\ 
  \multirow{2}*{$\mathcal{A}_{M}$}&Zero-shot          &\red{$10^{-25}$} & \red{0.22}      &&\red{$10^{-24}$} & \red{0.22}       &&\red{$10^{-3}$}    & $10^{-3}$          &&-&$10^{-2}$ & &\textbf{0.52} & $\mathbf{10^{-5}}$\\ 
  &Fine-tuning                                        &\red{$10^{-23}$} & \red{0.28}      &&\red{$10^{-27}$} & \red{0.28}       &&0.35& 0.01               &&-&$10^{-5}$ & &\textbf{0.50} & $\mathbf{10^{-6}}$\\ 
  \multirow{2}*{$\mathcal{A}_{Q}$}&Label-query        &\red{$10^{-27}$} & \red{0.20}      &&\red{$10^{-30}$} & \red{0.34}       &&\red{$10^{-5}$}    &\red{0.62}                &&-&$10^{-3}$ & &\textbf{0.52} & $\mathbf{10^{-4}}$\\ 
  &Logit-query                                        &\red{$10^{-27}$} & \red{0.23}      &&\red{$10^{-23}$} & \red{0.33}       &&\red{$10^{-6}$}    &\red{0.64}                && - &$10^{-3}$ & &\textbf{0.54} & $\mathbf{10^{-4}}$ \\ \hline
  Benign&Independent                                  &$10^{-20}$ & 0.33      &&$10^{-12}$ & 0.99       &&$10^{-22}$   &0.68                &&-&\red{$10^{-31}$} & &\textbf{0.00}  & \textbf{1.00} \\ 
  \bottomrule
  \end{tabular}
  }
\label{table:white-cifar10} 
\vspace{-0.3em}
\end{table*}

\begin{table*}[!t]
  \centering
  \caption{Results of defenses on the ImageNet dataset under the white-box setting. The best results among all defenses are indicated in boldface while the failed verification cases are marked in red.}
      \vspace{-0.8em}
  \scalebox{0.81}{
  \begin{tabular}{llcccccccccccccc}  
  \toprule  
  \multicolumn{2}{c}{\multirow{2}*{Model Stealing}} 
  &\multicolumn{2}{c}{BadNets}& &\multicolumn{2}{c}{Gradient Matching}& &\multicolumn{2}{c}{Entangled Watermarks}& &\multicolumn{2}{c}{Dataset Inference}& & \multicolumn{2}{c}{MOVE (Ours)}\\
  \cline{3-4}\cline{6-7}\cline{9-10}\cline{12-13}\cline{15-16}
  & &$\Delta \mu$ & p-value & & $\Delta \mu$ & p-value & & $\Delta \mu$ & p-value & & $\Delta \mu$ & p-value & & $\Delta \mu$ & p-value\\
  \hline
  $\mathcal{A}_{F}$ & Direct-copy                                     &0.87         &$10^{-10}$   &&0.77       &$10^{-10}$ &&\textbf{0.99} &$\mathbf{10^{-25}}$ &&-&$10^{-6}$&&0.90            &$10^{-5}$\\ 
  $\mathcal{A}_{D}$ &Distillation                     & \red{$10^{-4}$}    & \red{0.43}         && \red{$10^{-12}$} & \red{0.43}       && \red{$10^{-6}$}     & \red{0.19}                &&-&$10^{-3}$&&\textbf{0.61}    &$\mathbf{10^{-5}}$\\
  \multirow{2}*{$\mathcal{A}_{M}$}&Zero-shot & \red{$10^{-12}$}   & \red{0.33}         && \red{$10^{-18}$} & \red{0.43}       && \red{$10^{-3}$}     & \red{0.46}                &&-& $10^{-3}$&&\textbf{0.53}    &$\mathbf{10^{-4}}$\\
  &Fine-tuning                                        & \red{$10^{-20}$}   & \red{0.20}         &&\red{ $10^{-12}$} & \red{0.47}       && 0.46          &0.01                &&-&$10^{-4}$&&\textbf{0.60}    &$\mathbf{10^{-5}}$\\
  \multirow{2}*{$\mathcal{A}_{Q}$}&Label-query        & \red{$10^{-23}$}   & \red{0.29}         && \red{$10^{-22}$} & \red{0.50}       && \red{$10^{-7}$}     & \red{0.45}                &&-&$\bm{10^{-3}}$&&\textbf{0.55}    &$\mathbf{10^{-3}}$\\
  &Logit-query                                        & \red{$10^{-23}$}   & \red{0.38}         && \red{$10^{-12}$} & \red{0.22}       && \red{$10^{-6}$}     & \red{0.36}                &&-&$10^{-3}$&&\textbf{0.55}    &$\mathbf{10^{-4}}$\\ \hline
  Benign&Independent                                  &$10^{-24}$   &0.38         &&$10^{-23}$ &0.78       &&$\mathbf{10^{-30}}$    &0.55                &&-&\red{$10^{-10}$}     &&$10^{-5}$&\textbf{0.99}\\
  \bottomrule
  \end{tabular}
  }
  \label{table:white-imagenet}
    \vspace{-0.3em}
\end{table*}

\begin{table*}[!t]
  \centering
  \caption{\revision{Results of classical defenses} on the CIFAR-10 dataset under the black-box setting. The best results among all defenses are indicated in boldface while the failed verification cases are marked in red.}
        \vspace{-0.8em}
  \scalebox{0.81}{
  \begin{tabular}{llcccccccccccccc}  
  \toprule  
  \multicolumn{2}{c}{\multirow{2}*{Model Stealing}} 
  &\multicolumn{2}{c}{BadNets}& &\multicolumn{2}{c}{Gradient Matching}& &\multicolumn{2}{c}{Entangled Watermarks}& &\multicolumn{2}{c}{Dataset Inference}& & \multicolumn{2}{c}{MOVE (Ours)}\\
  \cline{3-4}\cline{6-7}\cline{9-10}\cline{12-13}\cline{15-16}
  & &$\Delta \mu$ & p-value & & $\Delta \mu$ & p-value & & $\Delta \mu$ & p-value & & $\Delta \mu$ & p-value & & $\Delta \mu$ & p-value\\
  \hline
  $\mathcal{A}_{F}$ &  Direct-copy                                 &0.79       &$10^{-60}$&&\red{0.03}      &$10^{-5}$&&0.54      &$10^{-35}$&&-&$10^{-39}$&&\textbf{0.84}& $\mathbf{10^{-74}}$  \\ 
  $\mathcal{A}_{D}$ &Distillation                &\red{$-10^{-3}$} & 0.02&&\red{0.01}&\red{ 0.14}&&\red{0.01}&$10^{-3}$ &&-&\red{ 0.32}&&\textbf{0.54}& $\mathbf{10^{-24}}$  \\
  \multirow{2}*{$\mathcal{A}_{M}$}&Zero-shot     &\red{$-10^{-25}$}&\red{0.29}&&\red{$10^{-4}$} &$10^{-3}$&&\red{$10^{-3}$}&\red{0.17}&&-&$10^{-5}$ &&\textbf{0.39}& $\mathbf{10^{-15}}$  \\ 
  &Fine-tuning                                   &\red{$10^{-3}$}&\red{0.08}&&\red{ $10^{-3}$}&\red{0.17}&&\red{$10^{-3}$}&\red{ 0.05}&&-&$10^{-6}$ &&\textbf{0.37}& $\mathbf{10^{-14}}$  \\
  \multirow{2}*{$\mathcal{A}_{Q}$}&Label-query   &\red{$10^{-4}$}&\red{0.11}&&\red{0.01}     &\red{0.11}&&\red{$10^{-3}$}&\red{0.05}&&-&$\mathbf{10^{-7}}$&&\textbf{0.07}& $10^{-3}$   \\ 
  &Logit-query                                   &\red{$10^{-3}$}&\red{0.10}&&\red{ $10^{-23}$}&\red{0.08}&&\red{$10^{-35}$}&\red{ 0.11}&&-&$10^{-3}$&&\textbf{0.17}& $\mathbf{10^{-6}}$  \\ \hline
  Benign&Independent                             &$10^{-20}$ &0.33&&$10^{-12}$&1.00&&\textbf{0.00}&\textbf{1.00}&&-&\red{$10^{-38}$} && $10^{-4}$ & 0.98  \\
  \bottomrule
  \end{tabular}
  }
  \label{table:black-cifar10} 
    \vspace{-0.8em}
\end{table*}

\vspace{0.3em}
\noindent \emph{Settings for Model Stealing.} Following the settings in \cite{maini2021dataset}, we conduct model stealing attacks illustrated in Section \ref{sec:model-stealing} to evaluate the effectiveness of different defenses. Besides, we also provide the results of examining a suspicious model which is not stolen from the victim (dubbed `Independent') for reference. Specifically, we implement the model distillation (dubbed `Distillation') \cite{hinton2015distilling} based on its open-sourced codes$\footnote{\url{https://github.com/thaonguyen19/ModelDistillation-PyTorch}}$. The stolen model is trained with SGD optimizer and an initial learning rate of 0.1, momentum of 0.9, and weight decay of $10^{-4}$; The zero-shot learning-based data-free distillation (dubbed `Zero-shot') \cite{fang2019data} is implemented based on the open-source codes\footnote{\url{https://github.com/VainF/Data-Free-Adversarial-Distillation}}. The stealing process is performed for 200 epochs with SGD optimizer and a learning rate of 0.1, momentum of 0.9, weight decay of $5\times 10^{-4}$, and batch size of 256; For the fine-tuning, the adversaries obtain stolen models by fine-tuning victim models on different datasets. Following the settings in \cite{maini2021dataset}, we randomly select 500,000 samples from the original TinyImages \cite{birhane2021large} as the substitute data to fine-tune the victim model for experiments on CIFAR-10. For the ImageNet experiments, we randomly choose samples with other 20 classes from the original ImageNet as the substitute data. We fine-tune the victim model for 5 epochs to obtain the stolen model; For the label-query attack (dubbed 'Label-query'), we train the stolen model for 20 epochs with a substitute dataset labeled by the victim model; For the logit-query attack (dubbed 'Logit-query'), we train the stolen model by minimizing the KL-divergence between its outputs ($i.e.$, logits) and those of the victim model for 20 epochs.

\vspace{0.3em}
\noindent \revision{\emph{Model Selection.} 
Following the settings of \cite{maini2021dataset}, we consider practical scenarios where the structure of the stolen model may be different from that of the victim model. Specifically, on CIFAR-10, the victim model is a WideResNet \cite{zagoruyko2016wide} with a depth of 28 and a widening factor of 10 (\ie, WRN-28-10). Except for direct-copy and fine-tuning (having the same structure), the stolen model is a smaller WRN-16-1; The victim model is ResNet-34 \cite{he2016deep} on ImageNet and the stolen model is ResNet-18 (except for direct-copy and fine-tuning).}

% 这部分拆分到6.2 和 6.3
%\vspace{0.3em}
%\noindent \emph{Settings for Defenses.} We compare our defense with dataset inference \cite{maini2021dataset} and model watermarking \cite{adi2018turning} with BadNets \cite{gu2019badnets}, gradient matching \cite{geiping2021witches}, and entangled watermarks \cite{jia2021entangled}. We poison 10\% training samples for all defenses. Besides, we adopt a white square in the lower right corner as the trigger pattern for BadNets and adopt an oil paint as the style image for our defense. Other settings are the same as those used in their original paper. We implement BadNets based on BackdoorBox \cite{li2022backdoorbox} and other methods based on their official open-sourced codes. An example of images ($e.g.$, poisoned images and the style image) involved in different defenses is shown in Figure \ref{fig:images}. In particular, similar to our method, dataset inference uses different methods under different settings, while other baseline defenses are designed under the black-box setting. Since methods designed under the black-box setting can also be used in the white-box scenarios, we also compare our MOVE with them under the white-box setting. 

\vspace{0.3em}
\noindent \emph{Evaluation Metrics.} We use the confidence score $\Delta \mu$ and p-value as the metric for our evaluation. Both of them are calculated based on the hypothesis-test with 10 sampled images under the white-box setting and 100 images under the black-box setting. In particular, except for the independent sources (which should not be regarded as stolen), the smaller the p-value and the larger the $\Delta \mu$, the better the defense. For the independent ones, the larger the p-value and the smaller the $\Delta \mu$, the better the method. Among all defenses, the best result is indicated in boldface and the failed verification cases are marked in red. \revision{We note that some baseline methods (\eg, BadNets) did not use a meta-classifier for ownership verification. In general, they verified ownership by determining whether the probability of the suspicious model in predicting watermarked images on a particular label (\ie, target label) is significantly greater than that of their original version on that label. In this case, the $\Delta \mu$ and p-value is calculated based on the alternative hypothesis $H_1: \mu_{S} > \mu_{B}$, where $\mu_{S} \triangleq f(X')_{y_t}$, $\mu_{B} \triangleq f(X)_{y_t}$, $f: \mathcal{X} \rightarrow [0,1]^K$ denotes the suspicious model, $y_t$ is the target label, $X'$ and $X$ denote the variable of watermarked image and that its benign version, respectively.}

\subsection{Results of MOVE under the White-box Setting}
\label{sec:results_white_box}

\noindent \revision{\emph{Settings.}} We compare our defense with dataset inference \cite{maini2021dataset} and model watermarking \cite{adi2018turning} with BadNets \cite{gu2019badnets}, gradient matching \cite{geiping2021witches}, and entangled watermarks \cite{jia2021entangled}. We poison 10\% training samples for all defenses. Besides, we adopt a white square in the lower right corner as the trigger pattern for BadNets and adopt an oil paint as the style image for our defense. Other settings are the same as those used in their original paper. We implement BadNets based on BackdoorBox \cite{li2022backdoorbox} and other methods based on their official open-source codes. An example of images ($e.g.$, poisoned images and the style image) involved in different defenses is shown in Figure \ref{fig:images}. In particular, similar to our method, dataset inference uses different methods under different settings, while other baseline defenses are designed under the black-box setting. Since methods designed under the black-box setting can also be used in the white-box scenarios, we also compare our MOVE with them under the white-box setting.

\vspace{0.3em}
\noindent \revision{\emph{Results.}} As shown in Table \ref{table:white-cifar10}-\ref{table:white-imagenet}, MOVE achieves the best performance in almost all cases under the white-box setting. For example, the p-value of our method is three orders of magnitude smaller than that of the dataset inference and six orders of magnitude smaller than that of the model watermarking in defending against the distillation-based model stealing on the CIFAR-10 dataset. The only exceptions appear when defending against the fully-accessible attack. In these cases, entangled watermark-based model watermarking has some advantages. Nevertheless, our method can still easily make correct predictions in these cases. In particular, our MOVE defense is the only method that can effectively identify whether there is model stealing in all cases. Other defenses either fail under many complicated attacks ($e.g.$, query-only attacks) or misjudge when there is no stealing. Besides, our defense has minor adverse effects on the performance of victim models. The accuracy of the model trained on benign CIFAR-10 and its transformed version is 91.79\% and 91.99\%, respectively; The accuracy of the model trained on benign ImageNet and its transformed version is 82.40\% and 80.40\%, respectively. This is mainly because we do not change the label of transformed images, and therefore, the transformation can be treated as data augmentation, which is mostly harmless.

\begin{table*}[!t]
  \centering
  \caption{\revision{Results of classical defenses} on the ImageNet dataset under the black-box setting. The best results among all defenses are indicated in boldface, while the failed verification cases are marked in red.}
        \vspace{-0.8em}
  \scalebox{0.81}{
  \begin{tabular}{llcccccccccccccc}  
  \toprule  
  \multicolumn{2}{c}{\multirow{2}*{Model Stealing}} 
  &\multicolumn{2}{c}{BadNets}& &\multicolumn{2}{c}{Gradient Matching}& &\multicolumn{2}{c}{Entangled Watermarks}& &\multicolumn{2}{c}{Dataset Inference}& & \multicolumn{2}{c}{MOVE (Ours)}\\
  \cline{3-4}\cline{6-7}\cline{9-10}\cline{12-13}\cline{15-16}
  & &$\Delta \mu$ & p-value & & $\Delta \mu$ & p-value & & $\Delta \mu$ & p-value & & $\Delta \mu$ & p-value & & $\Delta \mu$ & p-value\\
  \hline
  $\mathcal{A}_{F}$&  Direct-copy                            &0.87      &$10^{-10}$&&0.32      &$10^{-35}$&&\textbf{0.99}      &$10^{-48}$&&-&$10^{-65}$&&0.91&$\mathbf{10^{-135}}$\\
  $\mathcal{A}_{D}$ &Distillation             &\red{$10^{-4}$}&\red{0.09}&&\red{ $10^{-4}$}&\red{0.31}&&\red{$10^{-5}$}&\red{ 0.06}&&-&$10^{-50}$ &&\textbf{0.84}& $\mathbf{10^{-84}}$\\
  \multirow{2}*{$\mathcal{A}_{M}$}&Zero-shot  & \red{0.01}      &$10^{-4}$ &&\red{$10^{-3}$} &\red{ 0.30}&&\red{$10^{-3}$}&\red{0.33}&&-&$10^{-65}$ &&\textbf{0.88}&$\mathbf{10^{-113}}$ \\
  &Fine-tuning                                &\red{0.01}      &$10^{-4}$ &&\red{$10^{-12}$}&\red{0.21}&&\red{0.01}&\red{0.19}&&-&$10^{-55}$ &&\textbf{0.74}&$\mathbf{10^{-62}}$\\
  \multirow{2}*{$\mathcal{A}_{Q}$}&Label-query&\red{$10{-3}$}&\red{0.26}&&\red{$10^{-3}$}&\red{ 0.09}&&\red{$10^{-3}$}&\red{0.27}      &&-&$\mathbf{10^{-50}}$ &&\textbf{0.43}& $10^{-12}$\\
  &Logit-query                                &\red{$10^{-3}$}&\red{0.11}&&\red{$10^{-12}$}&\red{0.15}&&\red{$10^{-3}$}&\red{0.24}&&-&$\mathbf{10^{-54}}$ &&\textbf{0.61}& $10^{-37}$\\ \hline
  Benign&Independent                          &$10^{-24}$&0.38      &&$10^{-5}$ &0.28      &&$\mathbf{10^{-30}}$&\textbf{0.99}  &&-&\red{$10^{-50}$}&& $10^{-4}$ & 0.97 \\
  \bottomrule
  \end{tabular}
  }
  \label{table:black-imagenet}
\end{table*}

\begin{table*}[!t]
  \centering
  \caption{\revision{Results of advanced defenses on the CIFAR-10 dataset under the black-box setting. The best results among all defenses are indicated in boldface, while the failed verification cases are marked in red.}}
        \vspace{-0.8em}
  \scalebox{0.80}{
  \begin{tabular}{llccccccccc|cccccc|ccc}  
  \toprule  
  \multicolumn{2}{c}{\multirow{2}*{Model Stealing}} 
  &\multicolumn{2}{c}{APP}& &\multicolumn{2}{c}{PTYNet}& &\multicolumn{2}{c}{UAE}& &\multicolumn{2}{c}{UAPs}& &\multicolumn{2}{c}{Metafinger}& & \multicolumn{2}{|c}{MOVE (Ours)}\\
  \cline{3-4}\cline{6-7}\cline{9-10}\cline{12-13}\cline{15-16} \cline{18-19}
  & &$\Delta \mu$ & p-value & & $\Delta \mu$ & p-value & & $\Delta \mu$ & p-value & & $\Delta \mu$ & p-value & & $\Delta \mu$ & p-value& & $\Delta \mu$ & p-value\\
  \hline
  $\mathcal{A}_{F}$ &  Direct-copy                                 &0.78&	$10^{-67}$&&	0.23 &	$10^{-44}$&&0.40&	$10^{-14}$&&	0.85&	$10^{-72}$&&0.01   &$10^{-4}$&&	\textbf{0.84}&$\mathbf{10^{-74}}$\\
  $\mathcal{A}_{D}$ &Distillation                &0.04&	$10^{-3}$&&	\red{$10^{-3}$} &	\red{0.15} &&\red{$10^{-4}$}&	\red{0.19}&&	$10^{-3}$&	$10^{-3}$&&\red{$10^{-4}$}   &\red{0.21}&&	\textbf{0.54}&$\mathbf{10^{-24}}$\\
  \multirow{2}*{$\mathcal{A}_{M}$}&Zero-shot    &\red{$10^{-3}$}&	\red{0.50}	&&\red{$10^{-3}$}&	\red{0.31}    &&\red{$10^{-3}$}&	\red{0.04}&&	\red{$10^{-3}$}&	\red{0.03} &&\red{$10^{-4}$}&	\red{0.01}&&	\textbf{0.39}&$\mathbf{10^{-15}}$\\
  &Fine-tuning                                   &\textbf{0.66}&	$\mathbf{10^{-37}}$&&	0.22 &	$10^{-9}$ &&0.33&	$10^{-12}$&&	$10^{-3}$&	$10^{-3}$ &&$10^{-4}$&$10^{-3}$&&	0.37&$10^{-14}$\\
  \multirow{2}*{$\mathcal{A}_{Q}$}&Label-query   &\red{0.02}&	\red{0.02}    &&	\red{0.01} &	\red{0.27}    &&\red{$10^{-5}$}&	\red{0.23}&&	\red{$10^{-3}$}&	\red{0.04}   &&\red{$10^{-4}$}&	\red{0.23}&&	\textbf{0.07}&$\mathbf{10^{-3}}$\\
  &Logit-query                                   &\red{0.01}&	\red{0.05}	&&\red{$10^{-3}$}&	\red{0.12}    &&\red{$10^{-4}$}&	\red{0.14}&&	\red{$10^{-3}$}&	\red{0.06}   &&\red{$10^{-4}$} &	\red{0.09}&&	\textbf{0.17}&$\mathbf{10^{-6}}$\\ \hline
  Benign&Independent                             &0.02&	0.03	&&0.02	 &  1.00    &&$10^{-3}$&0.21  &&  $10^{-3}$&	0.05   &&\textbf{0.00}   &	\textbf{1.00}&&	$10^{-4}$&0.98 \\
  \bottomrule
  \end{tabular}
  }
  \label{table:new_baseline_cifar10} 
\end{table*}

\begin{table*}[!t]
  \centering
  \caption{\revision{Results of advanced defenses on the ImageNet dataset under the black-box setting. The best results among all defenses are indicated in boldface, while the failed verification cases are marked in red.}}
        \vspace{-0.8em}
  \scalebox{0.80}{
  \begin{tabular}{llccccccccc|cccccc|ccc}  
  \toprule  
  \multicolumn{2}{c}{\multirow{2}*{Model Stealing}} 
  &\multicolumn{2}{c}{APP}& &\multicolumn{2}{c}{PTYNet}& &\multicolumn{2}{c}{UAE}& &\multicolumn{2}{c}{UAPs}& &\multicolumn{2}{c}{Metafinger}& & \multicolumn{2}{c}{MOVE (Ours)}\\
  \cline{3-4}\cline{6-7}\cline{9-10}\cline{12-13}\cline{15-16} \cline{18-19}
  & &$\Delta \mu$ & p-value & & $\Delta \mu$ & p-value & & $\Delta \mu$ & p-value & & $\Delta \mu$ & p-value & & $\Delta \mu$ & p-value& & $\Delta \mu$ & p-value\\
  \hline
  $\mathcal{A}_{F}$ &  Direct-copy                                 &0.85     &$10^{-72}$&&0.01     &$10^{-4}$&&0.97&$10^{-7}$&&0.83     &$10^{-32}$&&0.08     &$10^{-3}$&&\textbf{0.91}&	$\mathbf{10^{-135}}$\\
  $\mathcal{A}_{D}$ &Distillation                &$10^{-3}$&$10^{-3}$ &&\red{$10^{-4}$}&\red{0.21}     &&0.53&$10^{-7}$&&\red{0.01}     &\red{0.11}      &&\red{$10^{-3}$}&\red{0.19}	&&\textbf{0.84}&	$\mathbf{10^{-84}}$\\
  \multirow{2}*{$\mathcal{A}_{M}$}&Zero-shot    &\red{$10^{-3}$}&\red{0.03}      &&\red{$10^{-4}$}&\red{0.01}     &&0.52&$10^{-5}$&&\red{$10^{-3}$}&\red{0.30}      &&\red{$10^{-4}$}&\red{0.08}	&&\textbf{0.88}&	$\mathbf{10^{-113}}$\\
  &Fine-tuning                                   &$10^{-3}$&$10^{-3}$ &&$10^{-4}$&$10^{-3}$&&0.50&$10^{-6}$&&\red{$10^{-3}$}&\red{0.24}      &&\red{$10^{-3}$}&\red{0.08}	&&\textbf{0.74}&	$\mathbf{10^{-62}}$\\
  \multirow{2}*{$\mathcal{A}_{Q}$}&Label-query   &\red{$10^{-3}$}&\red{0.04}      &&\red{$10^{-4}$}&\red{0.23}     &&0.52&$10^{-4}$&&\red{$10^{-3}$}&\red{0.21}      &&\red{$10^{-4}$}&\red{0.31}	&&\textbf{0.43}&	$\mathbf{10^{-12}}$\\
  &Logit-query                                   &\red{$10^{-3}$}&\red{0.06}      &&\red{$10^{-4}$}&\red{0.09}     &&0.54&$10^{-4}$&&\red{$10^{-3}$}&\red{0.23}      &&\red{$10^{-3}$}&\red{0.23}	&&\textbf{0.61}&$\mathbf{10^{-37}}$\\ \hline
  Benign&Independent                             &$10^{-3}$&0.05      &&0.00     &1.00     &&\textbf{0.00}&\textbf{1.00}     &&$10^{-3}$&0.14      &&\textbf{0.00}	   &\textbf{1.00}	&&$10^{-4}$&	0.97 \\
  \bottomrule
  \end{tabular}
  }
  \label{table:new_baseline_imgnet} 
\end{table*}

\subsection{Results of MOVE under the Black-box Setting}

\noindent \revision{\emph{Settings.} We compare our method with classical defense methods, including dataset inference \cite{maini2021dataset}, model watermarking \cite{adi2018turning} with BadNets \cite{gu2019badnets}, gradient matching \cite{geiping2021witches}, and entangled watermarks \cite{jia2021entangled}. For dataset inference, we use its black-box version in the experiments. The detailed settings for model watermarking with BadNets, gradient matching, and entangled watermarks are the same as those used in Section \ref{sec:results_white_box}. In addition to these classical methods, we also evaluate advanced defense methods as baselines, including both model watermark and model fingerprint. Specifically, for the model watermark, we compare our method with PTYNet \cite{wang2023free}, APP \cite{gan2023towards}, and UAE \cite{zhu2024reliable}. For model fingerprint, we compare our method with UAPs \cite{peng2022fingerprinting} and MetaFinger \cite{yang2022metafinger}. We implement these methods based on their official open-sourced codes with default settings.}

\vspace{0.3em}
\noindent \revision{\emph{Results.}} As shown in Table \ref{table:black-cifar10}-\ref{table:black-imagenet}, our MOVE defense still reaches promising performance under the black-box setting \revision{compared to classical black-box defenses (\eg, Dataset Inference).} For example, the p-value of our defense is over twenty and sixty orders of magnitude smaller than that of the dataset inference and entangled watermarks in defending against direct-copy on CIFAR-10 and ImageNet, respectively. In those cases where we do not get the best performance ($e.g.$, label-query and logit-query), our defense is usually the second-best, where it can still easily make correct predictions. \revision{As shown in Table \ref{table:new_baseline_cifar10}-\ref{table:new_baseline_imgnet}, MOVE consistently achieves the best performance in most cases over advanced defense methods. Although our method is the second-best in defending against fine-tuning attacks on the CIFAR-10 dataset, it remains highly effective without any failures.} In particular, same as the scenarios under the white-box setting, our MOVE defense is the only method that can effectively identify whether there is stealing in all cases. Other defenses either fail under many complicated stealing attacks or misjudge when there is no model stealing. These results verify the effectiveness of our defense again.

\subsection{Defending against Multi-Stage Model Stealing}
In previous experiments, the stolen model is obtained by a single stealing. In this section, we explore whether our method is still effective if there are multiple stealing stages.

\vspace{0.3em}
\noindent \emph{Settings.} We discuss two types of multi-stage stealing on the CIFAR-10 dataset, including stealing with the same attack and model structure and stealing with different attacks and model structures. In general, the first one is the easiest multi-stage attack, while the second one is the hardest. Other settings are the same as those in Section \ref{sec:exp_settings}.

\vspace{0.3em}
\noindent \emph{Results.} As shown in Table \ref{tab:multi-stage}, the p-value increases with the increase of the stage, which indicates that defending against multi-stage attacks is difficult. Nevertheless, the p-value $\leq 0.01$ in all cases under both white-box and black-box settings, $i.e.$, our method can identify the existence of model stealing even after multiple stealing stages. Besides, the p-value in defending the second type of multi-stage attack is significantly larger than that of the first one, showing that the second task is harder. %We will discuss how to better defend against the second type of attack in the future.

\begin{table}[!t]
\centering
\caption{Defending against multi-stage stealing.}
      \vspace{-0.8em}
\scalebox{0.83}{
\begin{tabular}{c|c|ccc}
\toprule
Setting$\downarrow$                    & Stage$\rightarrow$  & Stage 0      & Stage 1     & Stage 2   \\ \hline
\multirow{4}{*}{White-box} & Method$\rightarrow$ & Direct-copy & Zero-shot   & Zero-shot \\
                           & p-value$\rightarrow$      & $10^{-7}$   &     $10^{-5}$        &  $10^{-4}$         \\ \cline{2-5} 
                           & Method$\rightarrow$ & Direct-copy & Logit-query & Zero-shot \\
                           & p-value$\rightarrow$      & $10^{-7}$            & $10^{-4}$  &    0.01       \\ \hline \hline
\multirow{4}{*}{Black-box} & Method$\rightarrow$ & Direct-copy & Zero-shot   & Zero-shot \\
                           & p-value$\rightarrow$      & $10^{-74}$            &  $10^{-15}$           &     $10^{-5}$      \\ \cline{2-5} 
                           & Method$\rightarrow$ & Direct-copy & Logit-query & Zero-shot \\
                           & p-value$\rightarrow$      & $10^{-74}$            &  $10^{-6}$           &     $10^{-3}$      \\ \bottomrule
\end{tabular}
}
\vspace{-0.8em}
\label{tab:multi-stage}
\end{table}

\section{Discussion}
%In this section, we further explore the mechanisms and properties of our MOVE. Unless otherwise specified, all settings are the same as those in Section \ref{sec:exp}. 

\begin{figure*}[!t]
 \vspace{-0.5em}
    \centering
    \subfigure[MOVE under the White-box Setting]{
    \includegraphics[width=0.42\textwidth]{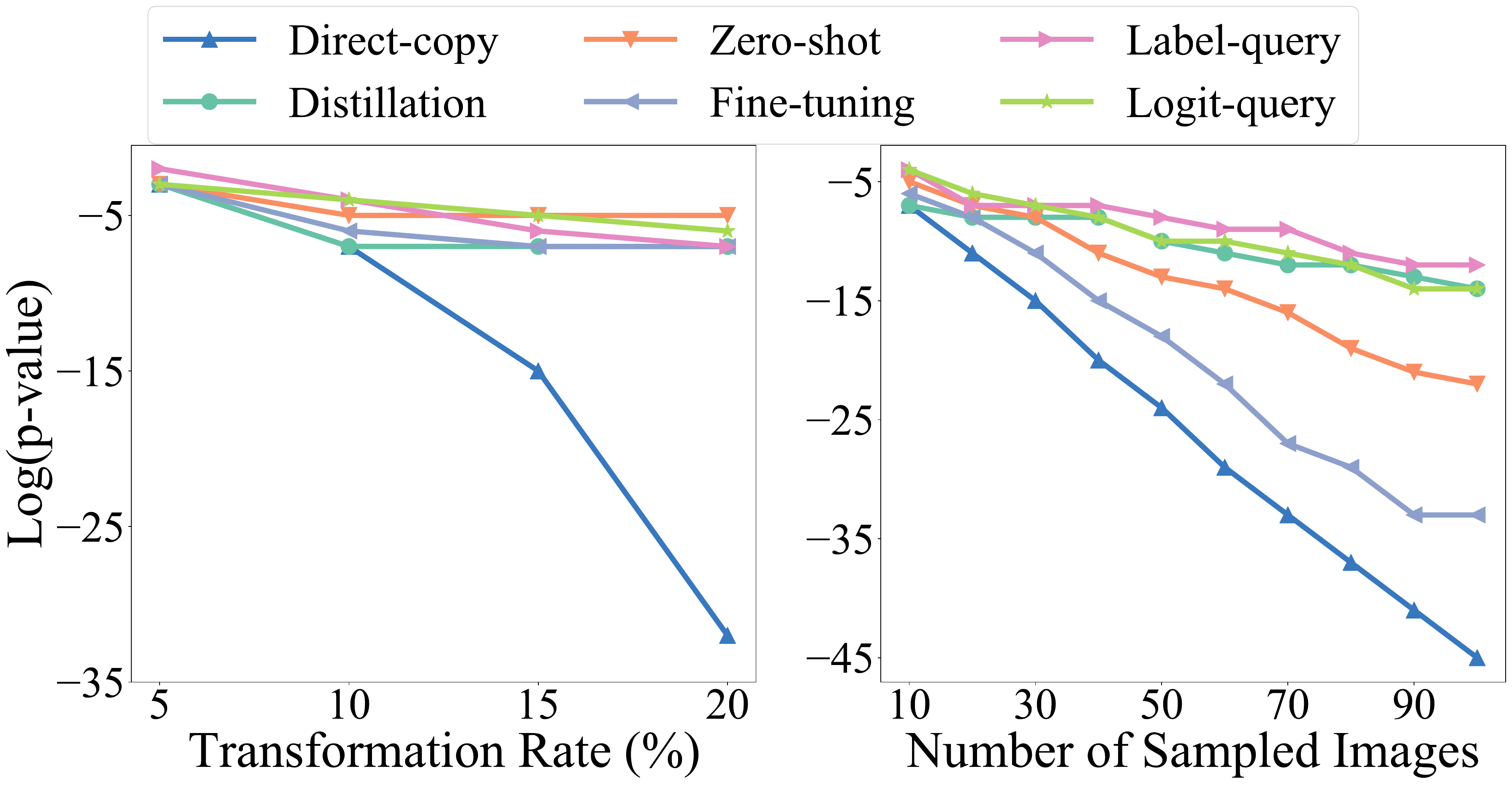}}
    \hspace{3em}
    \subfigure[MOVE under the Black-box Setting]{
    \includegraphics[width=0.42\textwidth]{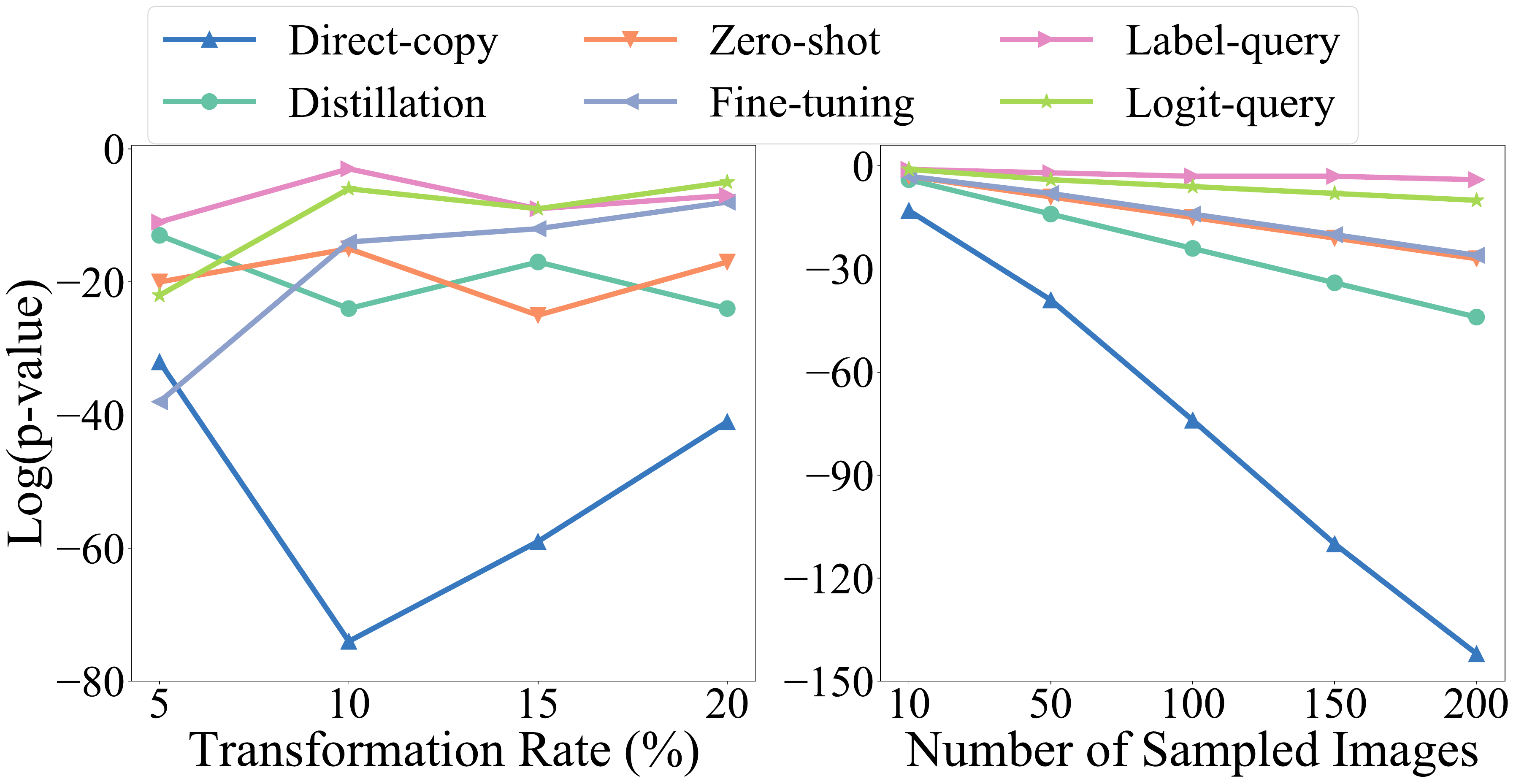}} 
    \vspace{-0.8em}
    \caption{The effects of the transformation rate (\%) and the number of sampled images of our MOVE on the CIFAR-10 dataset.}
    \label{fig:hyper}
\end{figure*}

\begin{table*}[!t]
\caption{The effectiveness of our defense with different style images on CIFAR-10 dataset.}
      \vspace{-0.8em}
\centering
\scalebox{0.9}{
\begin{tabular}{c|cccc|cccc}
\toprule
Method$\rightarrow$       & \multicolumn{4}{c|}{MOVE under the White-box Setting}                                                 & \multicolumn{4}{c}{MOVE under the Black-box Setting}                                                 \\ \hline
Pattern$\rightarrow$        & \multicolumn{2}{c|}{Pattern (a)}            & \multicolumn{2}{c|}{Pattern (b)} & \multicolumn{2}{c|}{Pattern (a)}            & \multicolumn{2}{c}{Pattern (b)} \\ \hline
Stealing Attack$\downarrow$, Metric$\rightarrow$ & $\Delta \mu$ & \multicolumn{1}{c|}{p-value} & $\Delta \mu$      & p-value      & $\Delta \mu$ & \multicolumn{1}{c|}{p-value} & $\Delta \mu$      & p-value     \\ \hline
Direct-copy    & 0.98         & \multicolumn{1}{c|}{$10^{-10}$}     & 0.98              & $10^{-12}$          & 0.92         & \multicolumn{1}{c|}{$10^{-126}$}    & 0.93              & $10^{-112}$        \\
Distillation   & 0.72         & \multicolumn{1}{c|}{$10^{-8}$}      & 0.63              & $10^{-7}$           & 0.71         & \multicolumn{1}{c|}{$10^{-41}$}     & 0.91              & $10^{-100}$        \\
Zero-shot      & 0.74         & \multicolumn{1}{c|}{$10^{-8}$}      & 0.67              & $10^{-7}$           & 0.67         & \multicolumn{1}{c|}{$10^{-37}$}     & 0.65              & $10^{-33}$        \\
Fine-tuning    & 0.21         & \multicolumn{1}{c|}{$10^{-7}$}      & 0.50              & $10^{-9}$           & 0.19         & \multicolumn{1}{c|}{$10^{-7}$}      & 0.21              & $10^{-9}$          \\
Label-query    & 0.68         & \multicolumn{1}{c|}{$10^{-8}$}      & 0.68              & $10^{-7}$           & 0.81         & \multicolumn{1}{c|}{$10^{-53}$}     & 0.74              & $10^{-41}$         \\
Logit-query    & 0.62         & \multicolumn{1}{c|}{$10^{-6}$}      & 0.73              & $10^{-7}$           & 0.6          & \multicolumn{1}{c|}{$10^{-28}$}     & 0.23              & $10^{-9}$          \\ \hline
Independent    & 0.00         & \multicolumn{1}{c|}{1.00}    & $10^{-9}$                & 0.99         & $10^{-4}$           & \multicolumn{1}{c|}{0.95}    & $10^{-4}$                & 0.96        \\ \bottomrule
\end{tabular}
}
\vspace{-0.5em}
\label{tab:effects_style}
\end{table*}

\begin{table*}[!t]
  \centering
  \caption{The verification performance of black-box MOVE with various victim backbones on CIFAR-10.}
  \vspace{-0.8em}
  \scalebox{0.85}{
  \begin{tabular}{llcccccccccccccc}  
  \toprule  
  \multicolumn{2}{c}{\multirow{2}*{}} 
  &\multicolumn{2}{c}{CLIP}& &\multicolumn{2}{c}{VGG-19}& &\multicolumn{2}{c}{AlexNet}& &\multicolumn{2}{c}{ResNet-18}& &\multicolumn{2}{c}{WideResNet}\\
  \cline{3-4}\cline{6-7}\cline{9-10}\cline{12-13}\cline{15-16}
  & &$\Delta \mu$ & p-value & & $\Delta \mu$ & p-value & & $\Delta \mu$ & p-value & & $\Delta \mu$ & p-value& & $\Delta \mu$ & p-value \\
  \hline
  $\mathcal{A}_{F}$ &  Direct-copy               &0.98&$10^{-14}$ && 0.78 & $10^{-50}$ &&0.29&$10^{-19}$&& 0.97 & $10^{-150}$ &&0.84& $10^{-74}$ \\ 
  $\mathcal{A}_{D}$ &Distillation                & 0.43& $10^{-5}$&& 0.10 & $10^{-3}$ &&0.09&$10^{-4}$ && 0.39& $10^{-21}$ &&0.54 & $10^{-24}$\\
  \multirow{2}*{$\mathcal{A}_{M}$}&Zero-shot     & 0.32 &$10^{-4}$ && 0.57 & $10^{-27}$ &&0.34&$10^{-27}$ &&0.90 & $10^{-50}$ &&0.39& $10^{-15}$ \\ 
  &Fine-tuning                                   & 0.87&$10^{-4}$  && 0.91 & $10^{-20}$ &&0.28 & $10^{-16}$&&0.94 &$10^{-66}$  &&0.37& $10^{-14}$ \\
  \multirow{2}*{$\mathcal{A}_{Q}$}&Label-query   &  0.68&$10^{-4}$ &&  0.15  &  $10^{-9}$ &&0.33&$10^{-26}$&& 0.36& $10^{-18}$ &&0.07& $10^{-3}$  \\ 
  &Logit-query                                   & 0.62&$10^{-3}$    &&  0.14 & $10^{-7}$ &&0.25&$10^{-16}$&&0.39 & $10^{-16}$ &&0.17& $10^{-6}$ \\ \hline
  Benign&Independent                             &0.00 &1.00&&0.00 & 1.00&&0.00&1.00&&0.00&1.00&&$10^{-4}$&0.98 \\
  \bottomrule
  \end{tabular}
  }
  \label{table:different_backbone_model} 
\end{table*}

\subsection{Effects of Key Hyper-parameters}
\label{sec:hyper}
%In this part, we discuss the effects of hyper-parameters and components involved in our method. 

\subsubsection{Effects of Transformation Rate}
In general, the larger the transformation rate $\gamma$, the more training samples are transformed during the training process of the victim model and therefore the `stronger' the external features. As shown in Figure \ref{fig:hyper}, as we expected, the p-value decreases with the increase of $\gamma$ in defending all stealing methods under both white-box. In other words, increasing the transformation rate can increase the performance of model verification. However, under the black-box setting, the changes in p-Value are relatively irregular. We speculate that this is probably because using prediction differences to approximately learn external features under the black-box setting has limited effects. Nevertheless, our method can successfully defend against all stealing attacks in all cases (p-value $\ll 0.01$). Besides, we need to emphasize that the increase of $\gamma$ may also lead to the accuracy decrease of victim models. Defenders should specify this hyper-parameter based on their specific requirements in practice.

\begin{figure}[!t]
    \centering
	\subfigure[]{
		\includegraphics[width=0.12\textwidth]{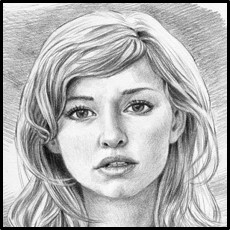}}
	\hspace{3em}
	\subfigure[]{
		\includegraphics[width=0.12\textwidth]{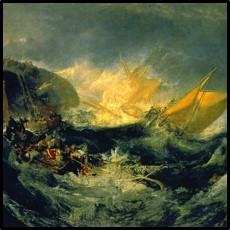}}
        \vspace{-0.5em}
    \caption{The new style images adopted for the evaluation.}
    \label{fig:styimg}
     \vspace{-0.8em}
\end{figure}

\subsubsection{Effects of the Number of Sampled Images}
Recall that our method needs to specify the number of sampled (transformed) images ($i.e.$, $m$) adopted in the hypothesis-based verification. In general, the larger the $m$, the less the adverse effects of the randomness involved in this process and therefore the more confident the verification. This is probably the main reason why the p-value also decreases with the increase of $m$, as shown in Figure \ref{fig:hyper}.

\subsubsection{Effects of Style Images}
In this part, we examine whether the proposed defense is still effective if we adopt other style images (as shown in Figure \ref{fig:styimg}). As shown in Table \ref{tab:effects_style}, the p-value $\ll 0.01$ in all cases under attack, while it is nearly 1 when there is no stealing, no matter under white-box or black-box setting. In other words, our method remains effective in defending against different stealing methods when different style images are used, although there will be some fluctuations in the results. We will further explore how to optimize the selection of style images in our future work.

\subsubsection{\revision{Effects of Model Structures}}

\revision{We hereby discuss whether our MOVE remains effective under different model structures. We consider two cases: \textbf{(1)} consistent stolen model but with different victim backbones and \textbf{(2)} inconsistent stolen model. We conduct experiments on CIFAR-10 using the more difficult black-box MOVE setting as examples for discussions. Specifically, for the first case, we exploit CLIP \cite{radford2021learning} (having a ViT-B/32 Transformer architecture as an image encoder), VGG \cite{simonyan2014very}, AlexNet \cite{krizhevsky2012imagenet}, and ResNet-18 \cite{he2016deep}. For the second case, we consider three victim-to-stolen model settings, including `VGG-19 to VGG-16', `ResNet-18 to VGG-16', and `WRN-28-10 to WRN-16-1'. As shown in Table \ref{table:different_backbone_model}-\ref{table:different_stolen_model}, our method is still highly effective under all settings, verifying its effectiveness and flexibility.}

\begin{table}[!t]
  \centering
  \caption{The verification performance of black-box MOVE with various victim-to-stolen cases on CIFAR-10.}
  \vspace{-0.8em}
  \scalebox{0.6}{
  \begin{tabular}{lccccccccc}  
  \toprule  
  \multirow{2}*{}&\multicolumn{2}{c}{VGG-19 to VGG-16}& &\multicolumn{2}{c}{ResNet-18 to VGG-16}&&\multicolumn{2}{c}{WRN-28-10 to WRN-16-1}\\
  \cline{2-3}\cline{5-6}\cline{8-9}
  & $\Delta \mu$ & p-value & & $\Delta \mu$ & p-value & & $\Delta \mu$ & p-value \\
  \hline
  Distillation  &0.08&$10^{-4}$  &&0.14&$10^{-6}$ &&0.54	&$10^{-24}$ \\
  Zero-shot     &0.61&$10^{-26}$ &&0.95& $10^{-5}$&&0.39	&$10^{-15}$   \\ 
  Label-query   &0.14&$10^{-12}$ &&0.70&$10^{-40}$&&0.07	&$10^{-3}$   \\ 
  Logit-query   &0.15&$10^{-6}$  &&0.65&$10^{-32}$&&0.17	&$10^{-6}$  \\
  \bottomrule
  \end{tabular}
  }
  \label{table:different_stolen_model} 
\end{table}

\begin{table}[!t]
  \caption{The effectiveness (p-value) of style transfer.}
  \centering
        \vspace{-0.8em}
  \scalebox{0.85}{
  \begin{tabular}{lccccc}  
  \toprule  
   & \multicolumn{2}{c}{CIFAR-10}& &\multicolumn{2}{c}{ImageNet} \\
  \cline{2-3}\cline{5-6}
  & \tabincell{c}{Patch-based\\Variant} & Ours& & \tabincell{c}{Patch-based\\Variant} & Ours\\\hline
  Direct-copy & $\mathbf{10^{-74}}$ & $10^{-7}$&& 0.01 &$\mathbf{10^{-5}}$\\ 
  Distillation & 0.17 & $\mathbf{10^{-7}}$ &&0.13 & $\mathbf{10^{-5}}$\\ 
  Zero-shot & 0.01 & $\mathbf{10^{-5}}$ && $10^{-3}$ & $\mathbf{10^{-4}}$\\ 
  Fine-tuning & $10^{-3}$ & $\mathbf{10^{-6}}$ &&$10^{-3}$ & $\mathbf{10^{-5}}$\\ 
  Label-query & $10^{-3}$ & $\mathbf{10^{-4}}$ &&0.02 & $\mathbf{10^{-3}}$\\ 
  Logit-query & $10^{-3}$ & $\mathbf{10^{-4}}$ &&0.01 & $\mathbf{10^{-4}}$\\
  \bottomrule
  \end{tabular}
  }
  \label{table:effectiveness_style}
  \vspace{-0.5em}
\end{table}

\subsection{The Ablation Study}
There are three key parts contained in our MOVE, including \textbf{1)} embedding external features with style transfer, \textbf{2)} using the sign vector of gradients instead of gradients themselves, and \textbf{3)} using meta-classifiers for verification. In this section, we verify their effectiveness. \revision{For simplicity, we use MOVE under the white-box setting for discussion, except for the data augmentation used only under the black-box setting.}

\subsubsection{The Effectiveness of Style Transfer}
\label{sec:eff_style}
To verify that the style watermark transfers better during the stealing process, we compare our method with its variant using the white-square patch (adopted in BadNets) to generate transformed images. As shown in Table \ref{table:effectiveness_style}, our method is significantly better than its patch-based variant. \revision{It is mostly because DNNs learn the texture information \cite{geirhos2019imagenet} better and the style watermark is bigger than the patch one.} This partly explains why our method works well.

\subsubsection{The Effectiveness of Meta-Classifier}
To verify that the meta-classifier is also useful, we compare the BadNets-based model watermarking with its extension, which also uses the meta-classifier (adopted in our MOVE defense) for ownership verification. In this case, the victim model is the backdoored one and the transformed image is the one containing backdoor triggers. As shown in Table \ref{table:effectiveness_meta}, adopting meta-classier significantly decrease the p-value in almost all cases, which verifies the effectiveness of the meta-classifier. \revision{It is mostly because meta-classifiers can learn complex features better}. These results also partly explains the effectiveness of our MOVE defense.

\begin{table}[!t]
  \caption{The effectiveness (p-value) of meta-classifier.}
  \centering
        \vspace{-0.8em}
  \scalebox{0.85}{
  \begin{tabular}{lccccc}  
  \toprule  
   & \multicolumn{2}{c}{CIFAR-10}& &\multicolumn{2}{c}{ImageNet} \\
  \cline{2-3}\cline{5-6}
  & w/o & w/ & & w/o & w/ \\\hline
  Direct-copy & $10^{-12}$&$\mathbf{10^{-37}}$&& $10^{-10}$ & $\mathbf{10^{-11}}$ \\ 
  Distillation & 0.32 & $\mathbf{10^{-3}}$ &&0.43 & $\mathbf{10^{-3}}$\\ 
  Zero-shot & 0.22 & $\mathbf{10^{-61}}$ && 0.33 & $\mathbf{10^{-3}}$\\ 
  Fine-tuning & 0.28 & $\mathbf{10^{-5}}$ &&0.20 & $\mathbf{10^{-13}}$\\ 
  Label-query & 0.20 & $\mathbf{10^{-50}}$ &&0.29 & $\mathbf{10^{-3}}$\\ 
  Logit-query & 0.23 & $\mathbf{10^{-3}}$ &&0.38 & $\mathbf{10^{-3}}$\\
  \bottomrule
  \end{tabular}
  }
  \label{table:effectiveness_meta} 
  \vspace{-0.8em}
\end{table}

\begin{table*}[!t]
\caption{The performance of our meta-classifier trained with different features.}
\centering
      \vspace{-0.8em}
\scalebox{0.88}{
\begin{tabular}{c|cc|cc|cccc}
\toprule
Dataset$\rightarrow$ & \multicolumn{4}{c|}{CIFAR-10} & \multicolumn{4}{c}{ImageNet}                                                                       \\ \hline
\multirow{2}{*}{Model Stealing$\downarrow$} & \multicolumn{2}{c|}{Gradient} & \multicolumn{2}{c|}{Sign of Gradient (Ours)} & \multicolumn{2}{c|}{Gradient} & \multicolumn{2}{c}{Sign of Gradient (Ours)} \\ \cline{2-9} 
& $\Delta \mu$ & p-value & $\Delta \mu$ & p-value & $\Delta \mu$ & \multicolumn{1}{c|}{p-value}& $\Delta \mu$ & p-value \\ \hline
Direct-copy       & 0.44 & $10^{-5}$ & $\bm{0.97}$ & $\bm{10^{-7}}$ &0.15      & \multicolumn{1}{c|}{$10^{-4}$} & $\bm{0.90}$ & $\bm{10^{-5}}$ \\ 
Distillation & 0.27 & 0.01      & $\bm{0.53}$ & $\bm{10^{-7}}$ &0.15      & \multicolumn{1}{c|}{$10^{-4}$} & $\bm{0.61}$ & $\bm{10^{-5}}$ \\
Zero-shot    & 0.03 & $10^{-3}$ & $\bm{0.52}$ & $\bm{10^{-5}}$ &0.12      & \multicolumn{1}{c|}{$10^{-3}$} & $\bm{0.53}$ & $\bm{10^{-4}}$ \\
Fine-tuning  & 0.04 & $10^{-5}$ & $\bm{0.50}$ & $\bm{10^{-6}}$ &0.13      & \multicolumn{1}{c|}{$10^{-3}$} & $\bm{0.60}$ & $\bm{10^{-5}}$ \\
Label-query  & 0.08 & $10^{-3}$ & $\bm{0.52}$ & $\bm{10^{-4}}$ &0.13      & \multicolumn{1}{c|}{$10^{-3}$} & $\bm{0.55}$ & $\bm{10^{-3}}$ \\
Logit-query  & 0.07 & $10^{-5}$ & $\bm{0.54}$ & $\bm{10^{-4}}$ &0.12      & \multicolumn{1}{c|}{$10^{-3}$} & $\bm{0.55}$ & $\bm{10^{-4}}$ \\ \hline
Independent & $\bm{0.00}$ & $\bm{1.00}$ & $\bm{0.00}$ & $\bm{1.00}$ &$\bm{10^{-10}}$ & \multicolumn{1}{c|}{$\bm{0.99}$} & $10^{-5}$ & $\bm{0.99}$ \\ \bottomrule
\end{tabular}
}
\label{table:signeffects} 
\end{table*}

\begin{table*}[!t]
  \centering
  \caption{The effects of data augmentation in black-box MOVE on CIFAR-10.}
        \vspace{-0.8em}
  \scalebox{0.9}{
  \begin{tabular}{llcccccccccccc}  
  \toprule  
  \multicolumn{2}{c}{\multirow{2}*{Model Stealing}} 
  &\multicolumn{2}{c}{No Augmentation}& &\multicolumn{2}{c}{Rotation}& &\multicolumn{2}{c}{Color Shifting}& &\multicolumn{2}{c}{Ours}\\
  \cline{3-4}\cline{6-7}\cline{9-10}\cline{12-13}
  & &$\Delta \mu$ & p-value & & $\Delta \mu$ & p-value & & $\Delta \mu$ & p-value & & $\Delta \mu$ & p-value \\
  \hline
  $\mathcal{A}_{F}$ &  Direct-copy               &0.77&$10^{-48}$&&0.37&$10^{-7}$&&	0.4	&$10^{-6}$&&	\textbf{0.84}&	$\mathbf{10^{-74}}$\\
  $\mathcal{A}_{D}$ &Distillation                &0.52&$10^{-4}$&&0.17	&$10^{-5}$	&&0.18	&$10^{-3}$&&	\textbf{0.54}&	$\mathbf{10^{-24}}$\\
  \multirow{2}*{$\mathcal{A}_{M}$}&Zero-shot     &\red{0.25}&\red{0.01}&&0.32	&$10^{-4}$	&&0.30	&$10^{-3}$	&&\textbf{0.39}	&$\mathbf{10^{-15}}$\\
  &Fine-tuning                                   &0.36&$10^{-3}$ & &0.35	&$10^{-5}$	&&0.35&	$10^{-4}$&&	\textbf{0.37}	&$\mathbf{10^{-14}}$\\
  \multirow{2}*{$\mathcal{A}_{Q}$}&Label-query   &\red{$10^{-3}$}&\red{0.19}& &\textbf{0.37}	&$\mathbf{10^{-4}}$	&&0.33	&$10^{-3}$&&	0.07&	$10^{-3}$\\ 
  &Logit-query                                   &\red{0.05}&\red{0.17}& &0.35	&$10^{-3}$	&&\textbf{0.36}	&$10^{-3}$	&&0.17	&$\mathbf{10^{-6}}$\\ \hline
  Benign&Independent                             &\textbf{0.00}&\textbf{1.00}&&\textbf{0.00}	&0.50	&&\textbf{0.00}	&0.50	&&$10^{-4}$	&0.98\\
  \bottomrule
  \end{tabular}
  }
  \label{table:aug} 
\vspace{-0.3em}
\end{table*}

\subsubsection{The Effectiveness of Sign Function}
In this section, we compare our method with its variant which directly adopts the gradients to train the meta-classifier. As shown in Table \ref{table:signeffects}, using the sign of gradients is significantly better than using gradients directly. This is probably because the `direction' of gradients contains more information compared with their `magnitude'. We will further explore it in the future.

\subsubsection{\revision{The Effectiveness of Data Augmentation}}

\revision{We hereby verify that data augmentation is critical for our MOVE under the black-box setting and discuss the effects of its transformation selection. We compare our method with its variants: \textbf{1)} MOVE without data augmentation (dubbed `No Augmentation'), \textbf{2)} MOVE with rotations (dubbed `Rotation'), and \textbf{3)} MOVE with color shifting (dubbed `Color Shifting'). Specifically, we exploit identical transformation, counterclockwise rotation with 20 and 45 degrees, and clockwise rotation with 20 and 45 degrees as five rotation-based transformations. We use identical transformation, random brightness adjustment, and three random hue jittering as five transformations in color shifting. As shown in Table \ref{table:aug}, without data augmentation, MOVE fails in many cases. In contrast, no matter under which augmentation, our method can accurately identify model stealing in all cases. Besides, our method is better than other variants in most cases. It is mostly because DNNs are vulnerable to rotation while color shifting leads to only minor changes.}

\subsection{Resistance to Potential Attacks}
\label{sec:resistance}
\revision{We hereby discuss the resistance of our defense to potential (adaptive) attacks}. We take MOVE under the white-box setting as an example to discuss for simplicity. \revision{Specifically, we discuss three representative backdoor defenses as classical potential attacks and design two adaptive attacks where adversaries know our MOVE method but without knowing the specific parameters used by model owners.}

\begin{figure*}[!t]
    \centering
    \subfigure[CIFAR-10]{
    \includegraphics[width=0.46\textwidth]{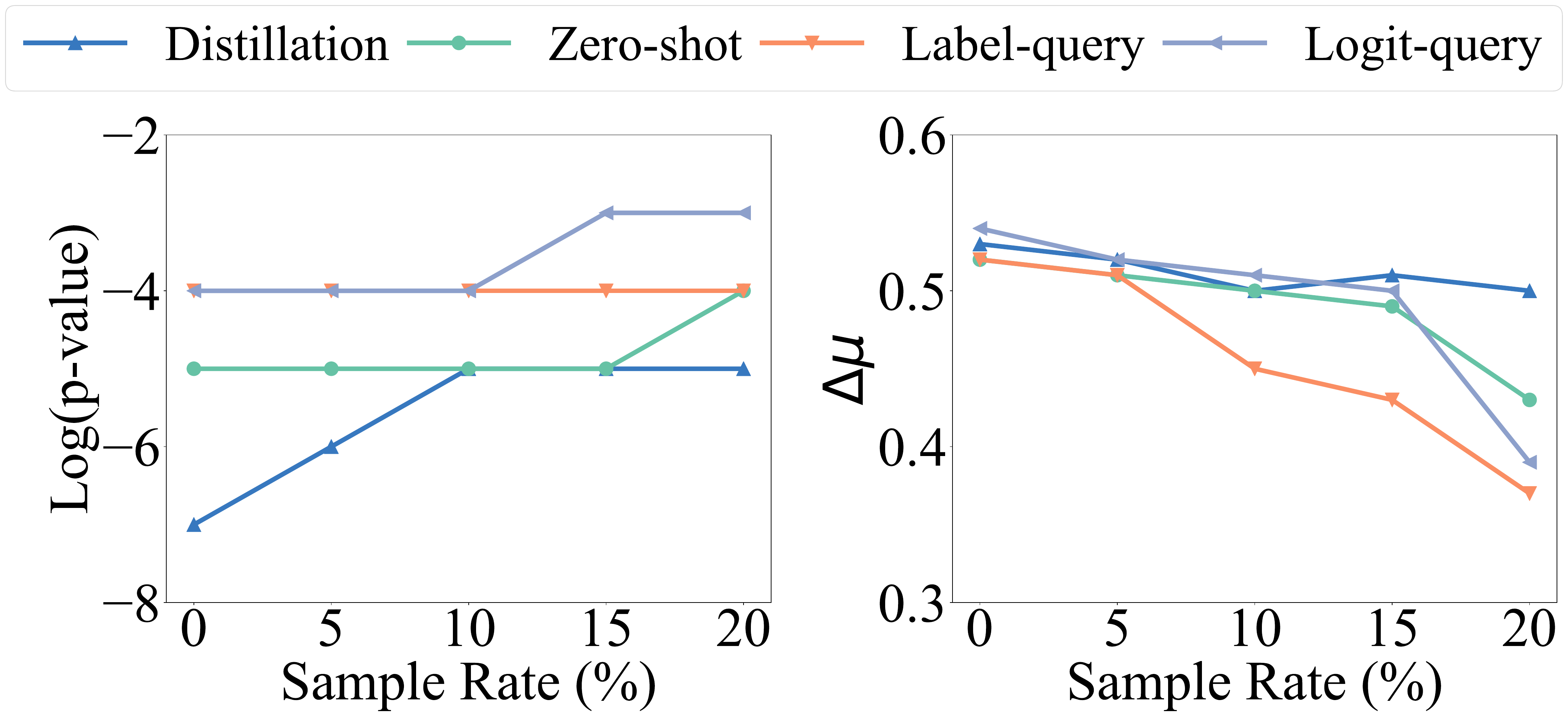}
    }\hspace{2em}
    \subfigure[ImageNet]{
    \includegraphics[width=0.46\textwidth]{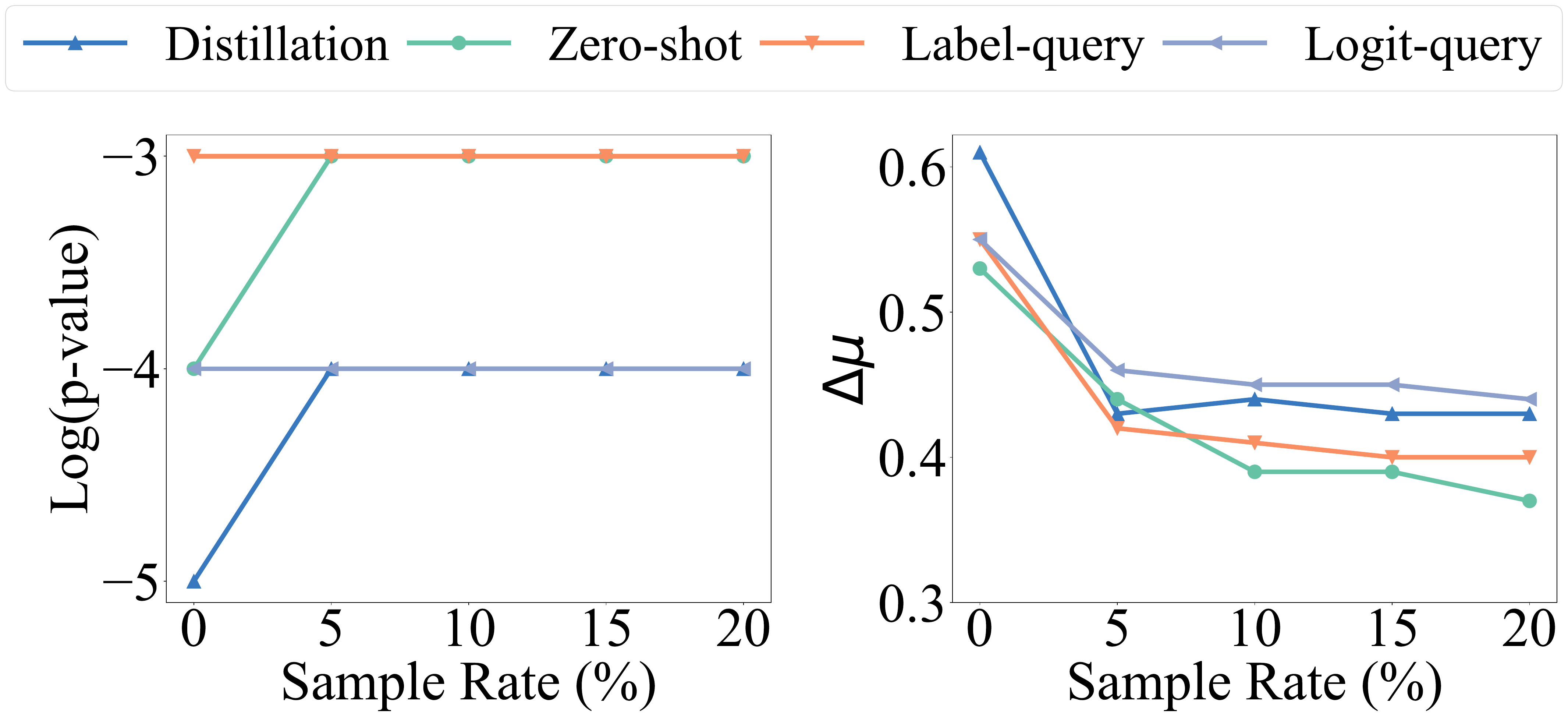}
    }
    \vspace{-0.8em}
    \caption{Resistance to the adaptive attack based on model fine-tuning with different sample rates.}
    \label{fig:resist_fine-tune}
    \vspace{-0.5em}
\end{figure*}

\subsubsection{Resistance to Model Fine-tuning}
In many cases, attackers have some benign local samples. As such, they may first fine-tuning the stolen model before deployment \cite{li2022backdoor,li2024purifying,li2024nearest}. This adaptive attack might be effective since the fine-tuning process may alleviate the effects of transformed images due to the catastrophic forgetting \cite{kirkpatrick2017overcoming} of DNNs. Specifically, we adopt $p\%$ benign training samples (where $p \in \{0, 5, 10, 15, 20\}$ is dubbed \emph{sample rate}) to fine-tune the whole stolen models generated by distillation, zero-shot, label-query, and logit-query. We also adopt the p-value and $\Delta \mu$ to measure the defense effectiveness.

As shown in Figure \ref{fig:resist_fine-tune}, the p-value increases while the $\Delta \mu$ decreases with the increase of sample rate. These results indicate that model fine-tuning has some benefits in reducing our defense effectiveness. However, our defense can still successfully detect all stealing attacks (p-value $<0.01$ and $\Delta \mu \gg 0$) even the sample rate is set to $20\%$. In other words, our defense is resistant to model fine-tuning. \revision{It is mostly because our method does not modify the original labels of watermarked images so that the watermarking task is more closely aligned with the main task.}

\begin{figure}[!t]
    \centering
    
    \subfigure[]{
		\includegraphics[width=0.11\textwidth]{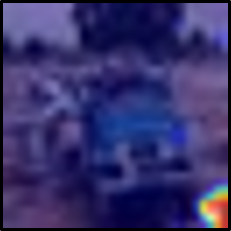}}
	%\hspace{2em}
	 \subfigure[]{
		\includegraphics[width=0.11\textwidth]{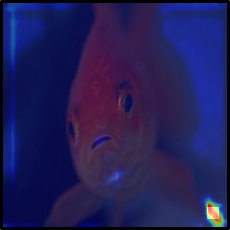}}
	%\hspace{2em}
	\subfigure[]{
		\includegraphics[width=0.11\textwidth]{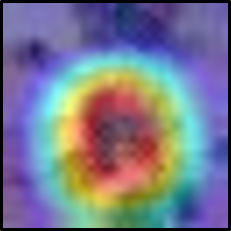}}
		%\hspace{2em}
	\subfigure[]{
		\includegraphics[width=0.11\textwidth]{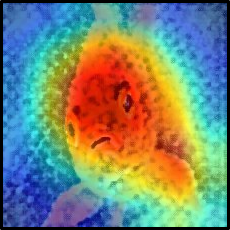}}
	\vspace{-0.8em}
    \caption{Salience maps of BadNets and our MOVE. \textbf{(a)-(b)}: salience maps of poisoned images generated by BadNets. \textbf{(c)-(d)}: salience maps of transformed images generated by MOVE. \textbf{(a)\&(c)}: salience maps of images from CIFAR-10. \textbf{(b)\&(d)}: salience maps of images from ImageNet.}
    \label{fig:saliancy}
    \vspace{-0.8em}
\end{figure}

\subsubsection{Resistance to Saliency-based Backdoor Detection}
These methods \cite{chou2020sentinet,doan2020februus,ya2024towards} use the saliency map to identify and remove potential trigger regions. Specifically, they first generate the saliency map of each sample and then calculate trigger regions based on the intersection of all saliency maps. As shown in Figure \ref{fig:saliancy}, the Grad-CAM mainly focuses on the trigger regions of images in BadNets, while it mainly focuses on the object outline towards images in our proposed defense. These results indicate that our MOVE is resistant to saliency-based backdoor detections.

\vspace{0.3em}
\subsubsection{Resistance to STRIP}
This method \cite{gao2021design} detects and filters poisoned samples based on the prediction randomness of samples generated by imposing various image patterns on the suspicious image. The randomness is measured by the entropy of the average prediction of those samples. \emph{The higher the entropy, the harder a method for STRIP to detect}. We compare the average entropy of all poisoned images in BadNets and that of all transformed images in our defense. As shown in Table \ref{tab:strip}, the entropy of our defense is significantly larger than that of the BadNets-based method. These results show that our defense is resistant to STRIP.

\begin{table}[!t]
\caption{Resistance (entropy) of BadNets-based model watermarking and our defense to STRIP. }
      \vspace{-0.8em}
\centering
\begin{tabular}{cc|cc}
\toprule
\multicolumn{2}{c|}{CIFAR-10} & \multicolumn{2}{c}{ImageNet} \\ \hline
BadNets & Ours & BadNets & Ours \\
0.01 & $\textbf{1.15}$ & 0.01 & $\textbf{0.77}$ \\ \bottomrule
\end{tabular}
\label{tab:strip}
\vspace{-0.8em}
\end{table}

\subsubsection{Resistance to Trigger Synthesis-based Detection}
\begin{figure}[!t]
    \centering
    \subfigure[]{
		\includegraphics[width=0.11\textwidth]{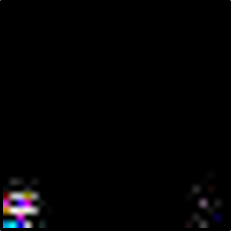}}
    \subfigure[]{
		\includegraphics[width=0.11\textwidth]{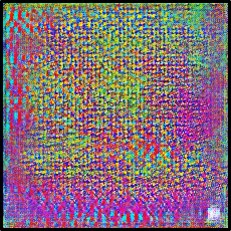}}
	\subfigure[]{
		\includegraphics[width=0.11\textwidth]{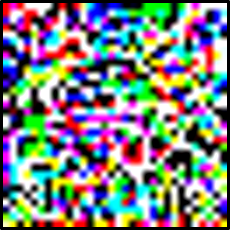}}
	\subfigure[]{
		\includegraphics[width=0.11\textwidth]{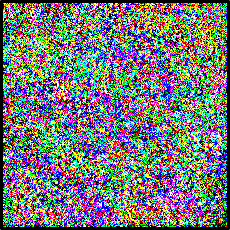}}
	\vspace{-0.8em}
    \caption{Reversed triggers. \textbf{First Row}: triggers of BadNets-based defense. \textbf{Second Row}: triggers of our method. \textbf{First Column}: triggers on CIFAR-10 dataset. \textbf{Second Column}: triggers on ImageNet dataset.}
    \label{fig:rev_triggers}
    \vspace{-0.8em}
\end{figure}

These methods \cite{wang2019neural,dong2021black,shen2021backdoor} detect poisoned images by reversing potential triggers contained in given suspicious DNNs. They have a latent assumption that the triggers should be sample-agnostic and the attack should be targeted. However, our defense does not satisfy these assumptions since the perturbations in our transformed images are sample-specific and we do not modify the label of those images. As shown in Figure \ref{fig:rev_triggers}, synthesized triggers of the BadNets-based method contain similar patterns to those used by defenders ($i.e.$, white-square on the bottom right corner) or its flipped version, whereas those of our method are meaningless. These results show that our defense is also resistant to these detection methods.

\subsubsection{\revision{Resistance to Overwriting Attack}}

\begin{figure}[!t]
    \centering
	\subfigure[]{
		\includegraphics[width=0.14\textwidth]{./fig/style.png}}
	\hspace{2em}
	\subfigure[]{
		\includegraphics[width=0.14\textwidth]{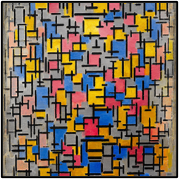}}
  \vspace{-0.8em}
    \caption{The style images used for the overwrite and style-removal attacks. \textbf{(a)} Style image used by defenders. \textbf{(b)} Style image used by adversaries.}
    \label{fig:overwrite_style}
\end{figure}

\revision{The adversaries may try to implant a different watermark (with a different style image) via fine-tuning to overwrite the implanted MOVE-based watermark in the victim model. We hereby evaluate the robustness of MOVE against this overwrite attack. Specifically, the adversaries first select a style image $\bm{\hat{x}_s}$ different from the original style image $\bm{x}_s$ and generate its transformed versions dataset $\hat{\mathcal{D}_t} = \{ (\bm{\hat{x}'}, y) | \bm{\hat{x}'} = T(\bm{x}, \bm{\hat{x}_s}), (\bm{x}, y) \in \hat{\mathcal{D}_s} \}$. The style images are shown in Figure \ref{fig:overwrite_style}. After that, the adversaries fine-tune the victim model, which has already learned the original style $\bm{x}_s$, for 50 epochs using this transformed dataset to overwrite the original watermark via}

\begin{equation}
\min_{\bm{\theta}} \sum_{(\bm{x}, y) \in \hat{\mathcal{D}_b} \cup \hat{\mathcal{D}_t}} \mathcal{L}(V_{\bm{\theta}}(\bm{x}), y),
\end{equation}
\revision{where $\hat{\mathcal{D}_b} \triangleq \mathcal{D} \backslash \hat{\mathcal{D}_s}$ and $\mathcal{L}(\cdot)$ is the loss function ($i.e.$, cross-entropy). As shown in Table \ref{table:adaptive_attack}, MOVE remains robust against the overwrite attack, consistently demonstrating its effectiveness without any failure cases. These results indicate that even when adversaries are aware of our defense method but lack specific parameter details, our approach effectively withstands adaptive attack strategies.}

\begin{table}[!t]
  \centering
  \caption{\revision{Resistance of our MOVE to overwrite attack (dubbed `Overwrite') and style-removal attack (dubbed `Style') on the CIFAR-10 dataset.}}
        \vspace{-0.8em}
  \scalebox{0.70}{
  \begin{tabular}{llccccccccc}  
  \toprule  
  \multicolumn{2}{c}{\multirow{2}*{Model Stealing}} 
  &\multicolumn{2}{c}{Overwrite}& &\multicolumn{2}{c}{Style}& &\multicolumn{2}{c}{No attack}\\
  \cline{3-4}\cline{6-7}\cline{9-10}
  & &$\Delta \mu$ & p-value & & $\Delta \mu$ & p-value & & $\Delta \mu$ & p-value \\
  \hline
  $\mathcal{A}_{F}$ &  Direct-copy               &0.29&$10^{-13}$ &&0.31&$10^{-42}$&&0.84	&$10^{-74}$\\
  $\mathcal{A}_{D}$ &Distillation                &0.05&$10^{-4}$&&0.06&$10^{-4}$ &&0.54	&$10^{-24}$\\
  \multirow{2}*{$\mathcal{A}_{M}$}&Zero-shot     &0.30&$10^{-3}$&&0.22&$10^{-5}$ &&	0.39	&$10^{-15}$\\
  &Fine-tuning                                   &0.10&$10^{-5}$&&0.11&$10^{-5}$ &&	0.37	&$10^{-14}$\\
  \multirow{2}*{$\mathcal{A}_{Q}$}&Label-query   &0.37&$10^{-3}$&&0.28&$10^{-3}$ &&	0.07&	$10^{-3}$\\
  &Logit-query                                   &0.25&$10^{-3}$&&0.26&$10^{-4}$ &&	0.17&	$10^{-6}$\\ \hline
  Benign&Independent                             &0.00&0.50	&&	0.00& 0.50	&&$10^{-4}$	&0.98\\
  \bottomrule
  \end{tabular}
  }
  \label{table:adaptive_attack} 
\end{table}

\subsubsection{\revision{Resistance to Style-removal Attack}}
\revision{The adversaries may try to remove the implanted MOVE-based watermark by minimizing the differences (\ie, cosine similarity) between the original images $\bm{x}$ and their watermarked versions (using a different style image $\bm{\hat{x}_s}$) through fine-tuning. We hereby evaluate the robustness of MOVE against this style-removal attack. The style images are shown in Figure \ref{fig:overwrite_style}. We fine-tune the watermarked victim model for 50 epochs to remove the watermark via}
\begin{equation}
    \min_{\bm{\theta}} \sum_{(\bm{x}, y) \in \mathcal{D}} \mathcal{L}(V_{\bm{\theta}}(\bm{x}), y) + \xi  \cdot \frac{V_{\bm{\theta}}(\bm{x}) \cdot V_{\bm{\theta}}(T(\bm{x}, \bm{\hat{x}_s}))}{\|V_{\bm{\theta}}(\bm{x})\| \|V_{\bm{\theta}}(T(\bm{x}, \bm{\hat{x}_s}))\|},
    \end{equation}
\revision{where $\mathcal{L}(\cdot)$ is the loss function ($i.e.$, cross-entropy) and $\xi$ is the corresponding hyper-parameters. We set $\xi=0.01$ in the experiment. As shown in Table \ref{table:adaptive_attack}, MOVE remains robust against the style-removal attack without any failure cases.}

\subsection{Relations with Related Methods}
\label{sec:relations}

\subsubsection{Relations with Membership Inference Attacks}
Membership inference attacks \cite{shokri2017membership} intend to identify whether a particular sample is used to train given DNNs. Similar to these attacks, our method also adopts some training samples for ownership verification. However, our defense aims to analyze whether the suspicious model contains the knowledge of external features rather than whether the model is trained on those transformed samples. To verify it, we design additional experiments, as follows:

\vspace{0.3em}
\noindent \emph{Settings.} For simplicity, we compare our method with its variant which adopts testing instead of training samples to generate transformed images used in ownership verification under the white-box setting. Except for that, other settings are the same as those stated in Section \ref{sec:exp_settings}.

\vspace{0.3em}
\noindent \emph{Results.} As shown in Table \ref{table:membership}, using testing images has similar effects to that of using training images in generating transformed images for ownership verification, although there are some small fluctuations. In other words, our method indeed examines the information of embedded external features rather than simply verifying whether the transformed images are used for training. These results verify that our defense is fundamentally different from membership inference attacks.

\begin{table*}[!t]
\caption{Results of ownership verification with transformed samples generated by training and testing samples.}
\vspace{-0.8em}
\centering
\scalebox{0.9}{
\begin{tabular}{c|cc|cc|cccc}
\toprule
Dataset$\rightarrow$ & \multicolumn{4}{c|}{CIFAR-10} & \multicolumn{4}{c}{ImageNet}                                                                       \\ \hline
\multirow{2}{*}{Model Stealing$\downarrow$} & \multicolumn{2}{c|}{Training Set} & \multicolumn{2}{c|}{Testing Set} & \multicolumn{2}{c|}{Training Set} & \multicolumn{2}{c}{Testing Set} \\ \cline{2-9} 
& $\Delta \mu$ & p-value & $\Delta \mu$ & p-value & $\Delta \mu$ & \multicolumn{1}{c|}{p-value}& $\Delta \mu$ & p-value \\ \hline
Direct-copy      & 0.97 & $10^{-7}$ & 0.96 & $10^{-7}$ & 0.90 & \multicolumn{1}{c|}{$10^{-5}$} & 0.93      & $10^{-7}$ \\
Distillation& 0.53 & $10^{-7}$ & 0.53 & $10^{-5}$ & 0.61 & \multicolumn{1}{c|}{$10^{-5}$} & 0.42      & $10^{-5}$ \\
Zero-shot   & 0.52 & $10^{-5}$ & 0.53 & $10^{-5}$ & 0.53 & \multicolumn{1}{c|}{$10^{-4}$} & 0.34      & $10^{-3}$ \\
Fine-tuning & 0.50 & $10^{-6}$ & 0.47 & $10^{-6}$ & 0.60 & \multicolumn{1}{c|}{$10^{-5}$} & 0.72      & $10^{-5}$ \\
Label-query & 0.52 & $10^{-4}$ & 0.52 & $10^{-4}$ & 0.55 & \multicolumn{1}{c|}{$10^{-3}$} & 0.40      & $10^{-3}$ \\
Logit-query & 0.54 & $10^{-4}$ & 0.53 & $10^{-4}$ & 0.55 & \multicolumn{1}{c|}{$10^{-4}$} & 0.48      & $10^{-4}$ \\ \hline
Independent & 0.00 & 1.00      & 0.00 & 1.00      & $10^{-5}$ & \multicolumn{1}{c|}{0.99} & $10^{-9}$ & 0.99 \\ \bottomrule
\end{tabular}
}
\vspace{-0.5em}
\label{table:membership}
\end{table*}

\subsubsection{Relations with Backdoor Attacks}
Similar to that of (poisoning-based) backdoor attacks \cite{qi2023revisiting,cai2024toward,gao2024backdoor}, our defense embeds pre-defined distinctive behaviors into DNNs via modifying some training samples. However, different from that of backdoor attacks, our method neither changes the label of poisoned samples nor only selects training samples with the specific category for poisoning. Accordingly, our defense will not introduce any hidden backdoor into the trained victim model. In other words, the dataset watermarking of our MOVE is also fundamentally different from backdoor attacks. This is probably the main reason why most existing backdoor defenses will have minor benefits in designing adaptive attacks against our defense, as illustrated in Section \ref{sec:resistance}.

\section{\revision{Potential Limitations and Future Work}}

\revision{We have to admit that it still has some potential limitations, although our MOVE method reaches promising performance under both white-box and black-box settings in defending against all stealing attacks.}

\revision{Firstly, our MOVE method directly exploits style transfer with a given defender-specified style image to implant external features. Although our experiments demonstrate that this approach is sufficiently effective, we believe that we may still use other approaches for feature injection or further improve our method by optimizing the style image. We will discuss them in our future work. }

\revision{Secondly, our method requires training a benign model and an external meta-classifier. This inevitably introduces some additional computational overhead. However, training a benign model for reference is a common practice in model watermarking and fingerprints to evaluate whether these techniques impact the model's performance \cite{jia2021entangled, maini2021dataset, wang2023free}. The training cost of a meta-classifier is mild since its structure is lightweight and it only needs limited training samples. For example, training it on CIFAR-10 and ImageNet requires 5 minutes using an NVIDIA A100 GPU. Nonetheless, we will discuss how to further improve the efficiency of our MOVE approach and design a benign model-free method in the future.}

\revision{Thirdly, this paper primarily focuses on defenses for image classification models. We will explore how to generalize our method in other modalities (\eg, text and audio) and tasks (\eg, text or image generation) in our future work.}

\section{Conclusion}
In this paper, we revisited the defenses against model stealing from the perspective of model ownership verification. We revealed that existing defenses suffered from low effectiveness and may even introduce additional security risks. Based on our analysis, we proposed a new effective and harmless model ownership verification ($i.e.$, MOVE), which examined whether the suspicious model contains the knowledge of defender-specified external features. We embedded external features by modifying a few training samples with style transfer without changing their label. In particular, we developed our MOVE defense under both white-box and black-box settings to provide comprehensive model protection. We evaluated our defense on both CIFAR-10 and ImageNet datasets. The experiments verified that our method can defend against various types of model stealing simultaneously while preserving high accuracy in predicting benign samples.

\section*{Acknowledgment}
This research is supported in part by the National Key Research and Development Program of China under Grant 2023YFB2904000, the National Natural Science Foundation of China under Grants (62171248, U23A20306, 62032021, and 62441619), and the Key Research Program of Hangzhou under grant 2024SZD1A27. This work was mostly done when Yiming Li was a Research Professor at the State Key Laboratory of Blockchain and Data Security, Zhejiang University, and a Ph.D. student at Tsinghua University. He is currently at Nanyang Technological University.

\bibliographystyle{IEEEtran}

\bibliography{reference}

\newpage

\appendix
\setcounter{theorem}{0}
\setcounter{equation}{0}

We hereby provide the proof of our Theorem \ref{thm1}. In general, it is based on the property of $t$-distribution.

\vspace{0.3em}
\begin{theorem}\label{thm1_app}
Given a (pre-trained) meta-classifier $C$ for distinguishing benign and stolen models, let $\beta_1 \triangleq \mathbb{P}(C(g_B)=1)$ and $\beta_2 \triangleq \mathbb{P}(C(g_V)=-1)$ denote its probability of Type-I and Type-II errors, respectively. Model owners can reject the previous null hypothesis $H_0$ at the significance level $\alpha$, if $\beta_1$ and $\beta_2$ satisfies 
$$
    \sqrt{m-1} \cdot (1-\beta_1 - \beta_2) - t_{\alpha} \cdot \sqrt{\beta_1 \cdot (1-\beta_1) + \beta_2 \cdot (1-\beta_2)}>0,
$$
where $t_{\alpha}$ is $\alpha$-quantile of t-distribution with $(m-1)$ degrees of freedom and $m$ is the number of verification samples.

\end{theorem}

% \begin{proof}

% Without loss of generality, let we take the white-box MOVE-based model ownership verification (\ie, Definition \ref{def:white-box}) as an example for the demonstration.

% Let $\bm{X}'$ is the variable of the transformed image, we have 

% \begin{gather}
% \beta_1 \triangleq \mathbb{P}(C(g_B)=1) = \mathbb{P}(C(g_B(\bm{X}'))=1), \\
% \beta_2 \triangleq \mathbb{P}(C(g_V)=-1) = \mathbb{P}(C(g_V(\bm{X}'))=-1),
% \end{gather}
% where $\beta_1$ is the probability of Type-I errors that the benign model is misclassified as a stolen model and $\beta_2$ is the probability of Type-II errors that the stolen model is misclassified as a non-stolen or independent model.

% We hereby seek to reject the original null hypothesis $H_0: \mu_{S} = \mu_{B} \ (H_1: \mu_{S} > \mu_{B})$ when the suspicious model $S$ is stolen from the victim. In this case, the original null hypothesis is equivalent to 
% \begin{equation}
% H_0': (1-\beta_2) = \beta_1 \ (H_1': (1-\beta_2) > \beta_1).
% \end{equation}

% Considering $m$ independent verification samples. Let $\hat{e}_1, \dots, \hat{e}_m$ denote the classification results of these samples and $E_1, \dots, E_m$ represent their classification events.

% For the benign model, the empirical Type-I error rate is:

% \begin{equation}
%   \beta_1 = \frac{1}{m} \sum_{i=1}^m E_i,  
% \end{equation}

% \end{proof}

\begin{proof}
Without loss of generality, let us take the white-box MOVE-based model ownership verification (\ie, Definition \ref{def:white-box}) as an example for the demonstration.

Given transformed image $\bm{X}'$, we have: 

\begin{gather}
\beta_1 \triangleq \mathbb{P}(C(g_B(\bm{X}')) = 1) \\
\beta_2 \triangleq \mathbb{P}(C(g_V(\bm{X}')) = -1),
\end{gather}
where $\beta_1$ is the probability of Type-I errors that the benign model is misclassified as a stolen model, and $\beta_2$ is the probability of Type-II errors that the stolen model is misclassified as a non-stolen or independent model.

We seek to reject the original null hypothesis $H_0: \mu_{S} = \mu_{B}$ (where $H_1: \mu_{S} > \mu_{B}$) when the suspicious model $S$ is stolen from the victim. In this case, the original null hypothesis is equivalent to 
\begin{equation}
H_0': (1 - \beta_2) = \beta_1 \quad (H_1': (1 - \beta_2) > \beta_1).
\end{equation}

Considering $m$ independent verification samples. Let $\hat{e}_1, \dots, \hat{e}_m$ denote their classification events.

For the benign model, the empirical Type-I error rate is:
\begin{equation}
\beta_1 = \frac{1}{m} \sum_{i=1}^m \hat{e}_i.
\end{equation}

For the stolen model, the empirical Type-II error rate is:
\begin{equation}
1 - \beta_2 = \frac{1}{m} \sum_{i=1}^m (1 - \hat{e}_i).
\end{equation}

Thus, we have the following Bernoulli distributions:
\begin{gather}
\beta_1 \sim \frac{1}{m} B(m, \beta_1), \\
1 - \beta_2  \sim \frac{1}{m} B(m, 1 - \beta_2).
\end{gather}

According to the Central Limit Theorem, when $m$ is sufficiently large, $\beta_1$ and $\beta_2$ follow Gaussian distribution:
\begin{gather}
\beta_1  \sim \mathcal{N}\left(\beta_1, \frac{\beta_1(1-\beta_1)}{m}\right), \\
1 - \beta_2 \sim \mathcal{N}\left(1 - \beta_2, \frac{\beta_2(1-\beta_2)}{m}\right).
\end{gather}

We define the difference $D$ between the success rate of classifying the stolen model correctly and the error rate for the benign model as:
\begin{equation}
D = (1 - \beta_2) - \beta_1.
\end{equation}
Due to the properties of the normal distribution, $D$ also follows a normal distribution:
\begin{equation}
D \sim \mathcal{N}\left((1 - \beta_2) - \beta_1, \frac{\beta_2(1 - \beta_2) + \beta_1(1 - \beta_1)}{m}\right).
\end{equation}

Next, we construct the $t$-statistic as follows:
\begin{equation}
T = \frac{\sqrt{m} \cdot D}{s},
\end{equation}
where $s$ is the standard deviation of $D$, and is given by:
\begin{equation}
s^2 = \frac{1}{m-1} \sum_{i=1}^m (D_i - \bar{D})^2.
\end{equation}

Under the null hypothesis $H_0'$, the $T$ statistic follows a $t$-distribution with $(m-1)$ degrees of freedom:
\begin{equation}
T \sim t(m-1).
\end{equation}

To reject $H_0'$ at significance level $\alpha$, we require:
\begin{equation}
T > t_\alpha,
\end{equation}
where $t_\alpha$ is the $\alpha$-quantile of the $t$-distribution.

Expanding the expression for $T$, we have:
\begin{equation} \label{eq:near-final}
\frac{\sqrt{m} \cdot [(1 - \beta_2) - \beta_1]}{\sqrt{\beta_2 \cdot (1-\beta_2) + \beta_1 \cdot (1-\beta_1)}} > t_\alpha.
\end{equation}

Rearranging the inequality (\ref{eq:near-final}), we finally have:
\begin{equation}
\sqrt{m-1} \cdot (1-\beta_1 - \beta_2) - t_{\alpha} \cdot \sqrt{\beta_1 \cdot (1-\beta_1) + \beta_2 \cdot (1-\beta_2)}>0.
\end{equation}

\end{proof}

\begin{IEEEbiography}[{\includegraphics[width=0.9in,height=1.1in,clip,keepaspectratio]{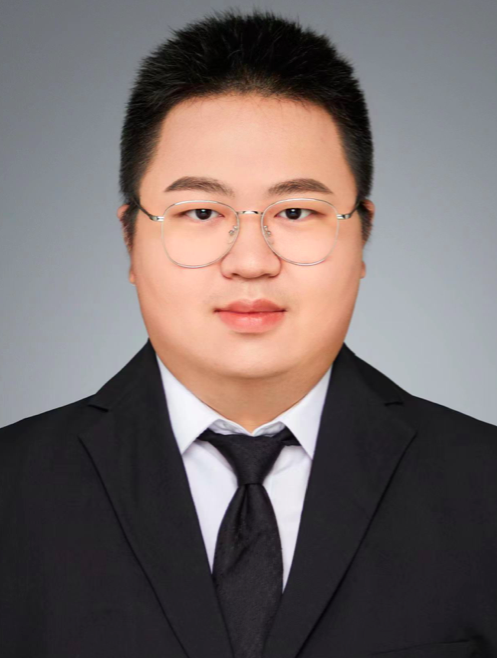}}]{Dr. Yiming Li} is currently a Research Fellow at Nanyang Technological University. Before that, he was a Research Professor with the Outstanding Junior Faculty Award in the State Key Laboratory of Blockchain and Data Security at Zhejiang University and also in HIC-ZJU. He received his Ph.D. degree with honors in Computer Science and Technology from Tsinghua University in 2023 and his B.S. degree with honors in Mathematics from Ningbo University in 2018. His research interests are in the domain of Trustworthy ML and Responsible AI, especially AI Risk Management and AI Copyright Protection. His research has been published in multiple top-tier conferences and journals, such as S\&P, USENIX Security, NDSS, ICML, NeurIPS, ICLR, CVPR, ICCV, TPAMI, and IEEE TIFS. He served as the Associate Editor of Pattern Recognition, the Area Chair of ICML and ACM MM, the Senior Program Committee Member of AAAI, and IJCAI, and the Reviewer of IEEE TPAMI, IJCV, IEEE TIFS, IEEE TDSC, etc. His research has been featured by major media outlets, such as IEEE Spectrum and MIT Technology Review. He was the recipient of the Best Paper Award at PAKDD and the Rising Star Award at WAIC.
\end{IEEEbiography}

\begin{IEEEbiography}[{\includegraphics[width=0.9in,height=1.1in,clip,keepaspectratio]{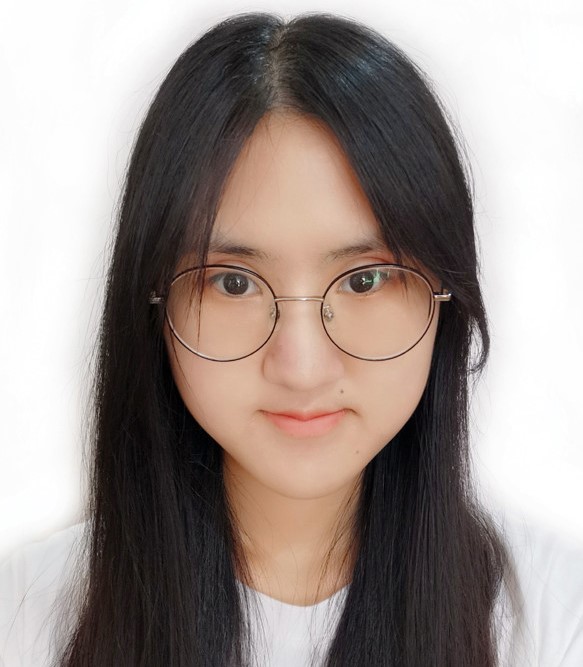}}]{Linghui Zhu} received the B.S. degree in Computer Science and Technology from Nankai University, Tianjin, China, in 2020. She is currently pursuing the master degree in Tsinghua Shenzhen International Graduate School, Tsinghua University. Her research interests are in the domain of data security and federated learning.
\end{IEEEbiography}

\begin{IEEEbiography}[{\includegraphics[width=0.9in,height=1.1in,clip,keepaspectratio]{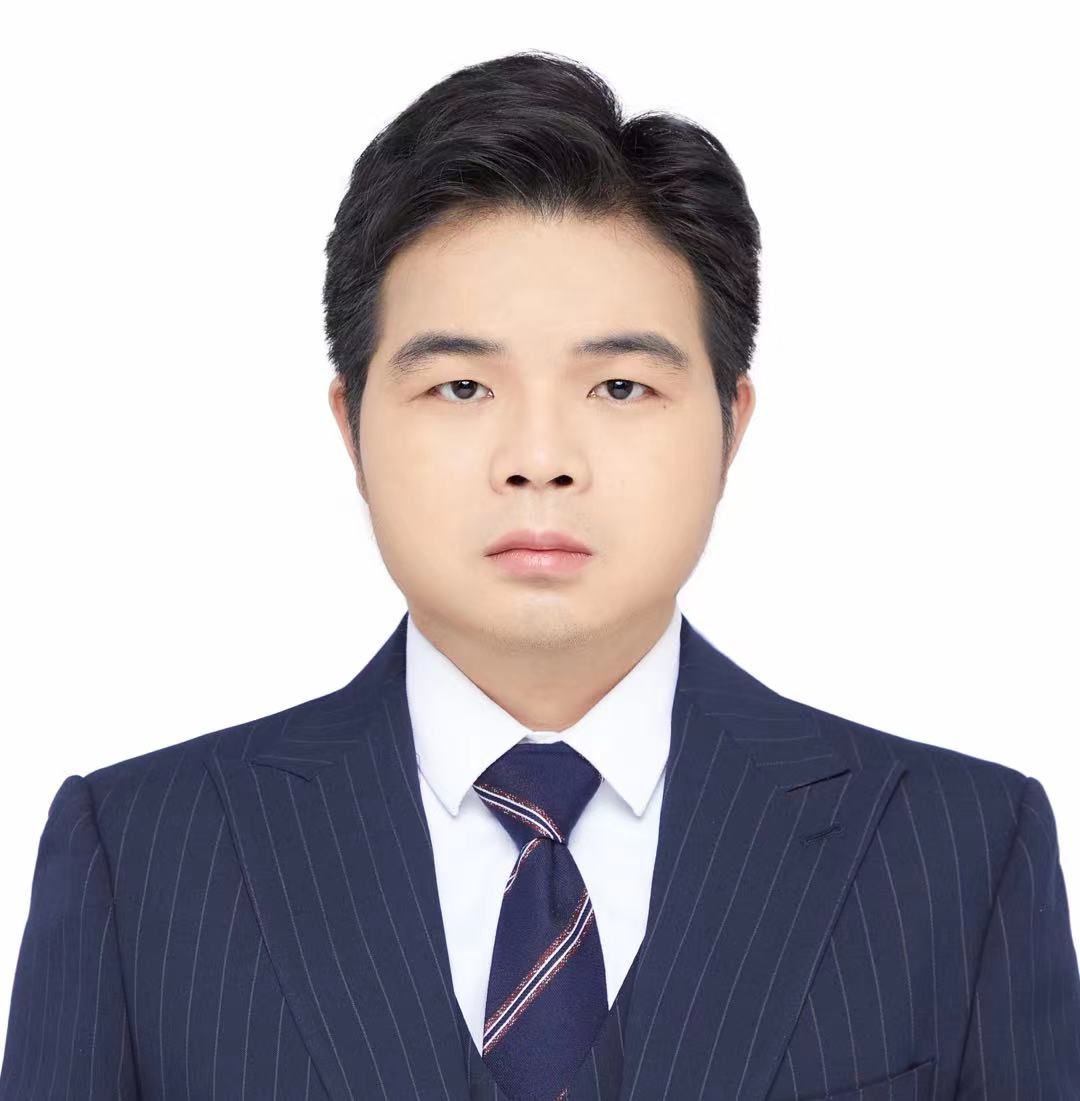}}]{Dr. Xiaojun Jia} received his Ph.D. degree in  State Key Laboratory of Information Security, Institute
of Information Engineering, Chinese Academy of Sciences and School of
Cyber Security, University of Chinese Academy of Sciences, Beijing. He is now a Research Fellow in Cyber Security Research Centre @ NTU, Nanyang Technological University, Singapore. His research interests include computer vision, deep learning and adversarial machine learning.
\end{IEEEbiography}

\begin{IEEEbiography}[{\includegraphics[width=0.9in,height=1.1in,clip,keepaspectratio]{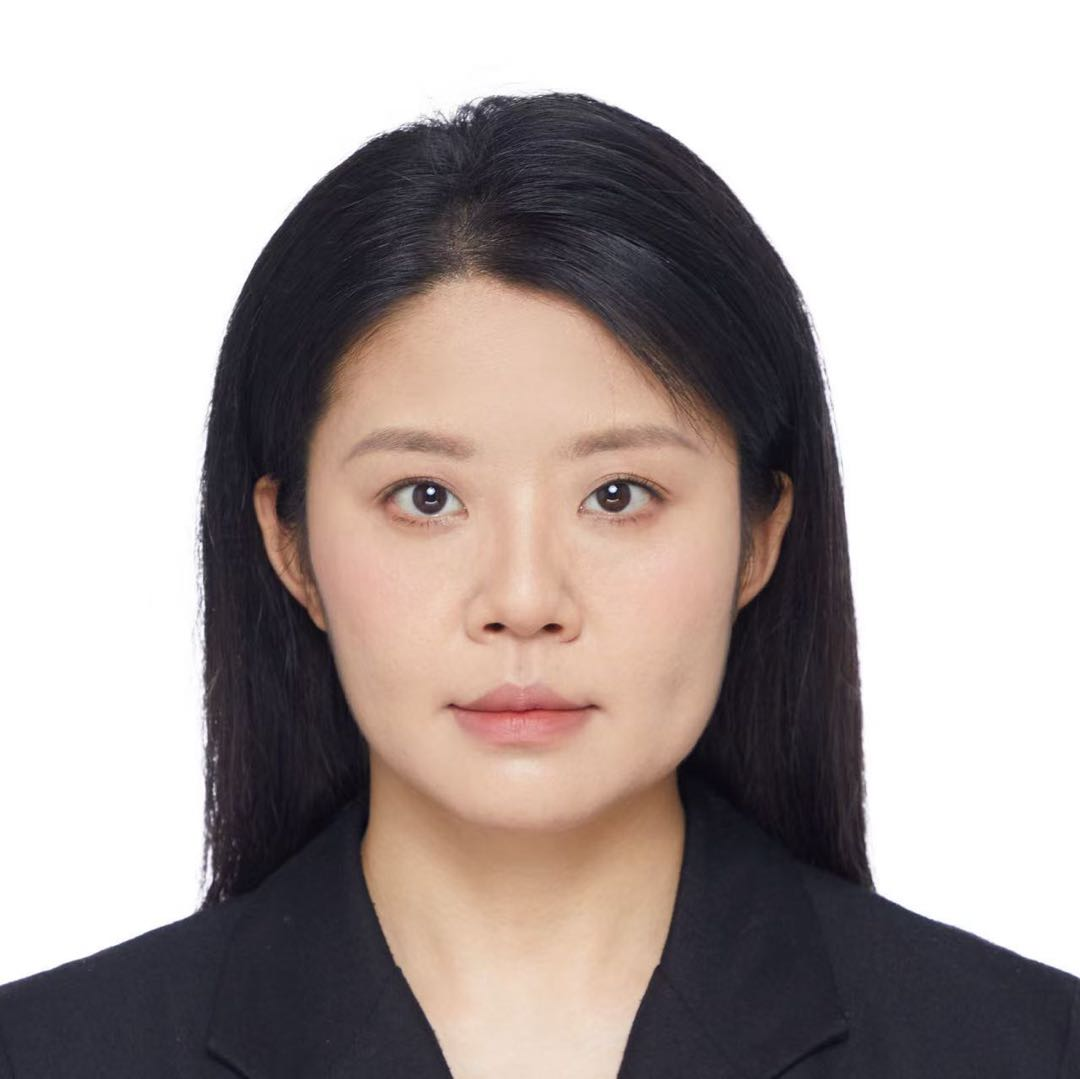}}]{Dr. Yang Bai} received the B.S. degree in Electronic Information Science and Technology and the Ph.D. degree in Computer Science and Technology from Tsinghua University in 2017 and 2022, respectively. She is now a senior research engineer at ByteDance Inc in the US. Her research interests include Robust Machine Learning and LLM Safety. Her research has been published in multiple top-tier conferences, including ICLR, NeurIPS, CVPR, ICCV, ECCV, etc. 
\end{IEEEbiography}

\begin{IEEEbiography}[{\includegraphics[width=0.9in,height=1.1in,clip,keepaspectratio]{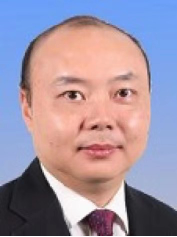}}]{Dr. Yong Jiang} received his M.S. and Ph.D. degrees in computer science from Tsinghua University, China, in 1998 and 2002, respectively. Since 2002, he has been with the Tsinghua Shenzhen International Graduate School of Tsinghua University, Guangdong, China, where he is currently a full professor. His research interests include computer vision, machine learning, Internet architecture and its protocols, IP routing technology, etc. He has received several best paper awards from top-tier conferences and his research has been published in multiple top-tier journals and conferences, including IEEE ToC, IEEE TMM, IEEE TSP, CVPR, ICLR, etc.
\end{IEEEbiography}

\begin{IEEEbiography}[{\includegraphics[width=0.9in,height=1.1in,clip,keepaspectratio]{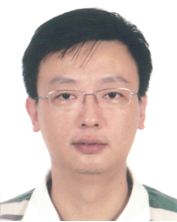}}]{Dr. Shu-Tao Xia} received the B.S. degree in mathematics and the Ph.D. degree in applied mathematics from Nankai University, Tianjin, China, in 1992 and 1997, respectively. Since January 2004, he has been with the Tsinghua Shenzhen International Graduate School of Tsinghua University, Guangdong, China, where he is currently a full professor. His research interests include coding and information theory, machine learning, and deep learning. His research has been published in multiple top-tier journals and conferences, including IEEE TPAMI, IEEE TIP, CVPR, ICLR, etc.
\end{IEEEbiography}

\begin{IEEEbiography}[{\includegraphics[width=0.9in,height=1.1in,clip,keepaspectratio]{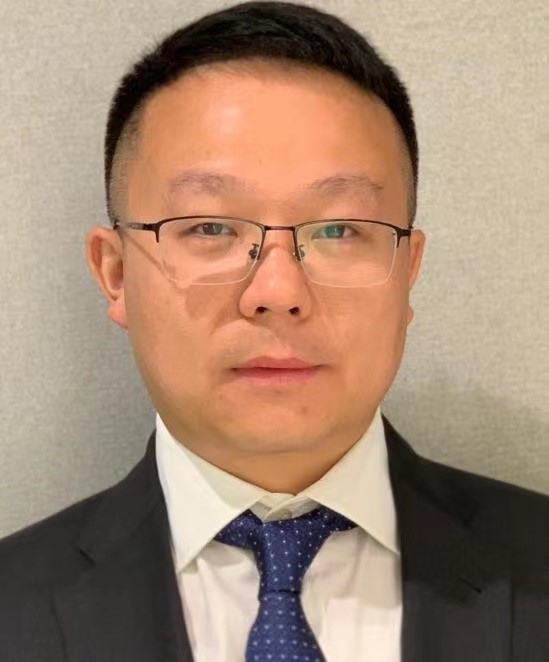}}]{Dr. Xiaochun Cao} received the B.S. and M.S. degrees in computer science from Beihang University, China, and the Ph.D. degree in computer science from the University of Central Florida, USA. He is with the School of Cyber Science and Technology, Sun Yat-sen University, China, as a Full Professor and the Dean. He has authored and co-authored multiple top-tier journal and conference papers. He is on the Editorial Boards of the IEEE TIP, IEEE TMM, IEEE TCSVT. He was the recipient of the Best Student Paper Award at the ICPR (2004, 2010). He is the Fellow of the IET.
\end{IEEEbiography}

\begin{IEEEbiography}[{\includegraphics[width=0.9in,height=1.1in,clip,keepaspectratio]{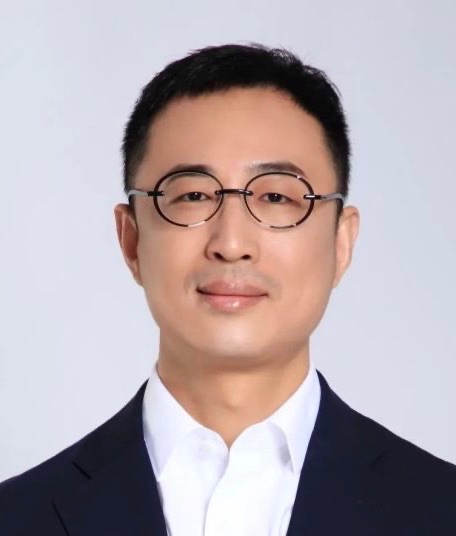}}]{Dr. Kui Ren} (Fellow, IEEE) received the Ph.D. degree in electrical and computer engineering from Worcester Polytechnic Institute. He is currently the Dean of the College of Computer Science and Technology, Zhejiang University, and the Executive Deputy Director of the State Key Laboratory of Blockchain and Data Security. Before that, he was a SUNY Empire Innovation Professor with the State University of New York at Buffalo, USA. His current research interests include data security, the IoT security, AI security, and privacy. He is a fellow of ACM. He is a member of ACM ASIACCS Steering Committee and S\&T Committee of the Ministry of Education of China. He received many recognitions, including the Guohua Distinguished Scholar Award of ZJU, the IEEE CISTC Technical Recognition Award, the SUNY Chancellor's Research Excellence Award, the Sigma Xi Research Excellence Award, and the NSF CAREER Award. He has published extensively in peer-reviewed journals and conferences and received the Test-of-Time Paper Award from IEEE INFOCOM and many Best Paper Awards, including ACM MobiSys, IEEE ICDCS, IEEE ICNP, IEEE GLOBECOM, and ACM/IEEE IWQoS. His H-index is 96, with a total citation exceeding 50,000 according to Google Scholar. He is a frequent reviewer of funding agencies internationally and serves on the editorial boards of many IEEE and ACM journals. Among others, he serves as the Chair of SIGSAC of the ACM China Council.
\end{IEEEbiography}

\end{document}